%% file: main.tex
\documentclass[sigconf]{acmart}

\input{lib}

\AtBeginDocument{%
  }

\copyrightyear{2026}
\acmYear{2026}
\setcopyright{cc}
\setcctype{by-nc-nd}
\acmConference[CHI '26]{Proceedings of the 2026 CHI Conference on Human Factors in Computing Systems}{April 13--17, 2026}{Barcelona, Spain}
\acmBooktitle{Proceedings of the 2026 CHI Conference on Human Factors in Computing Systems (CHI '26), April 13--17, 2026, Barcelona, Spain}
\acmPrice{}
\acmDOI{10.1145/3772318.3791342}
\acmISBN{979-8-4007-2278-3/2026/04}

\acmSubmissionID{8900}

\begin{document}

\title{\sysname{}: Bridging the Gap between EFL Learning and Work through AI-Generated Work-Related Quizzes}

\settopmatter{authorsperrow=5}

\author{Yeonsun Yang}
\authornote{Yeonsun Yang conducted this work as a research intern at NAVER AI Lab.}
\orcid{0009-0002-4653-8159}
\affiliation{%
  \institution{DGIST}
  \country{Republic of Korea}
}
\email{diddustjs98@dgist.ac.kr}

\author{Sang Won Lee}
\authornote{Sang Won Lee conducted this work as a visiting scholar at NAVER AI Lab.}
\orcid{0000-0002-1026-315X}
\affiliation{%
  \institution{Virginia Tech}
  \city{Blacksburg}
  \state{VA}
  \country{USA}
}
\email{sangwonlee@vt.edu}

\author{Jean Y. Song}
\orcid{0000-0003-4379-3971}
\affiliation{%
  \institution{Yonsei University}
  \country{Republic of Korea}
}
\email{jeansong@yonsei.ac.kr}

\author{Sangdoo Yun}
\orcid{0000-0002-0417-8450}
\affiliation{%
  \institution{NAVER AI Lab}
  \country{Republic of Korea}
}
\email{oodgnas@gmail.com}

\author{Young-Ho Kim}
\orcid{0000-0002-2681-2774}
\affiliation{%
  \institution{NAVER AI Lab}
  \country{Republic of Korea}
}
\email{yghokim@younghokim.net}

\begin{abstract}
\input{sections/00_abstract}
\end{abstract}



\begin{teaserfigure}
  \includegraphics[width=\textwidth]{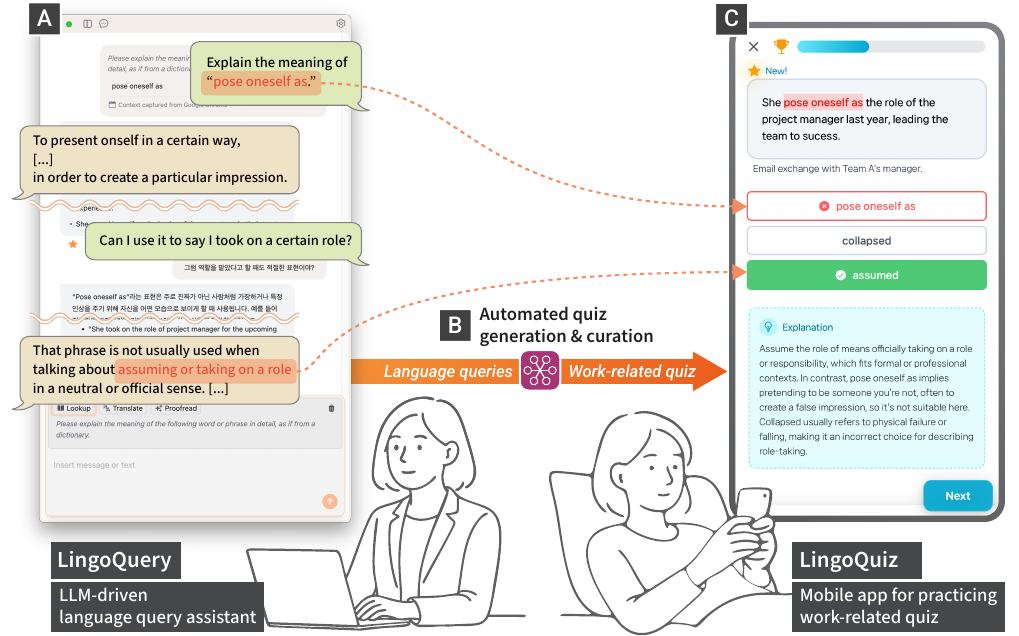}
  \caption{\sysname{} consists of three components. In \queryapp{} for desktop \blackrectsmall{A}, information workers can interact with an LLM-based chatbot for English-related language queries. The automated quiz generation pipeline \blackrectsmall{B} produces and curates multiple choice English questions using the query interactions of \queryapp{} as materials. In \quizapp{} \blackrectsmall{C} on a smartphone, workers can later review their language queries by completing the generated quizzes. (Please refer to our supplementary video, available at
\sourcecode{}, which demonstrates the interactions.)}
  \Description{A teaser illustration showing the workflow of LingoQ. On the left, a woman uses LingoQuery on her computer to chat with an AI assistant for English-related queries. In the center, her chat messages are automatically transformed into quiz questions. On the right, the same woman is lying on her bed and solving the generated quizzes on her smartphone using LingoQuiz.}
  \label{fig:teaser}
\end{teaserfigure}

\maketitle

\input{sections/01_introduction}
\input{sections/02_relatedwork}

\input{sections/03_formative}
\input{sections/04_system}
\input{sections/05_userstudy}
\input{sections/06_results}

\input{sections/07_discussion}

\input{sections/08_conclusion}

\begin{acks}
We thank our study participants from the formative studies, English test validation, expert evaluation, and the deployment study, for their time and efforts.
This work was supported through a research internship at NAVER AI Lab of NAVER Cloud, and in part by the National Research Foundation of Korea (NRF) grant funded by the Korea government (MSIT) (No. RS-2024-00353125).
\end{acks}

\bibliographystyle{ACM-Reference-Format}
\bibliography{bibliography}

\onecolumn
\appendix
\input{appendix/prompts}
\input{appendix/quizevaluation}

\input{appendix/expertrubric}


\end{document}

%% file: lib.tex
\usepackage{multirow}
\usepackage{color}
\usepackage{colortbl}
\usepackage{stfloats}
\usepackage{subcaption}
\usepackage{enumitem}
\usepackage{graphicx}
\usepackage{subcaption}
\usepackage{tabularx}
\usepackage{longfbox}
\usepackage{kotex}
\usepackage{xcolor}
\usepackage{fontawesome5}

\makeatletter
\newdimen\@tempdimd
\makeatother

\usepackage{listings}
\newfboxstyle{boxparam}{padding=1.5pt, padding-left=2pt, padding-right=2pt, height=7.5pt, background-color=lightgray, border-radius=2pt, border-right-width=1pt, border-bottom-width=1pt, border-color=gray}

\newcommand{\rectwrapsmall}[2]{\lfbox[boxparam, border-radius=0pt, padding-left=2pt, padding-right=2pt, height=5.5pt, border-width=0pt, background-color=#1]{\sffamily{\textcolor{white}{#2}}}}

\newcommand{\blackrectsmall}[1]{\rectwrapsmall{darkgray}{#1}}

\definecolor{quotebackground}{HTML}{EFEFEF}
\definecolor{tableheader}{HTML}{EFEFEF}
\definecolor{tablegrayline}{HTML}{e0e0e0}

\newcommand{\sysname}{\textsc{LingoQ}}
\newcommand{\queryapp}{\textsc{LingoQuery}}
\newcommand{\quizapp}{\textsc{LingoQuiz}}

\newcommand{\eg}{\textit{e.g.}}
\newcommand{\ie}{\textit{i.e.}}
\newcommand{\etal}{\textit{et al.}}
\newcommand{\cf}{\textit{c.f.}}

\newcommand{\labelphantom}[1]{%
  \parbox{0pt}{\phantomsubcaption\label{#1}}%
}

\newcommand{\starmark}{{\color{orange}\hspace{0.2em}\scalebox{0.9}{\faStar}\hspace{0.2em}}}

\newcommand{\shortcut}{{\color{darkgray}\hspace{0.2em}\scalebox{0.9}{\faKeyboard}\hspace{0.2em}}}

\newtoggle{clean}
\toggletrue{clean} 

\DeclareRobustCommand{\revised}[1]{%
  \iftoggle{clean}{#1}{\textcolor{blue}{#1}}%
}

\newcommand{\idx}[1]
{}

\newtoggle{cameraclean}
\toggletrue{cameraclean} 

\DeclareRobustCommand{\cameraready}[1]{%
  \iftoggle{cameraclean}{#1}{\textcolor{red}{#1}}%
}

\newenvironment{revisedenv}{%
    \begingroup
    \iftoggle{clean}{}{%
        \color{blue}%
    }%
    \ignorespaces
}{%
    \endgroup
    \ignorespacesafterend
}

\newcommand{\deleted}[1]{%
  \ifthenelse{\boolean{clean}}{}{%
    \textcolor{cyan}{\st{#1}}%
  }%
}

\newcommand{\deletedsubsection}[1]{%
  \ifthenelse{\boolean{clean}}{}{%
    \subsection{\textcolor{cyan}{[Deleted] #1}}%
  }%
}

\newcommand{\circledigit}[1]{\textbf{\normalsize{\textsf{\textcircled{\footnotesize{#1}}}}}}

\newcommand{\ipstart}[1]{\vspace{1mm} \noindent{\textbf{\textit{#1.}}}}

\newcommand{\sourcecode}{\urlstyle{tt}\url{https://naver-ai.github.io/lingo-q}}

%% file: sections/00_abstract.tex
Non-native English speakers performing English-related tasks at work struggle to sustain EFL learning, despite their motivation. Often, study materials are disconnected from their work context.
\revised{Our formative study revealed that reviewing work-related English becomes burdensome with current systems, especially after work.}
Although workers rely on LLM-based assistants to address their immediate needs, these interactions may not directly contribute to their English skills.
We present \sysname{}, an AI-mediated system that allows workers to practice English using quizzes generated from their LLM queries during work.
\sysname{} leverages these on-the-fly queries using AI to generate personalized quizzes that workers can review and practice on their smartphones. 
We conducted a three-week deployment study with 28 EFL workers to evaluate \sysname{}.
Participants valued the \cameraready{quality-assured, work-situated quizzes and} constantly engaging with the app during the study. 
This active engagement improved self-efficacy and led to learning gains for beginners and, potentially, for intermediate learners.
\revised{Drawing on these results, we discuss design implications for leveraging workers' growing reliance on LLMs to foster proficiency and engagement while respecting work boundaries and ethics.}


%% file: sections/01_introduction.tex
\section{Introduction}


In the global economy, where online resources are predominantly in English, information workers\footnote{In this work, we use the term \textit{information worker} to refer to workers whose primary job role involves gathering, synthesizing, and producing new information~\cite{kuhlthau1999role}. In this paper, we will use the term `workers' to specifically refer to information workers.} who are not native English speakers often need English proficiency for their jobs. This requirement includes understanding English texts used in papers, articles, and reports, as well as the ability to communicate via email.
%
To enhance their English proficiency, information workers frequently engage in self-directed language learning using mobile applications like Duolingo~\cite{duolingo}, Babbel~\cite{babbel}, Memrise~\cite{memrise}, and RosettaStone~\cite{rosettastone} that offers access to learning anytime, anywhere through mobile phones~\cite{viberg2012mobile}.
%
However, the mobile language learning apps often create study materials from generic situations like traveling or business meetings, limiting the depth and practical relevance of vocabulary and conversational skills. Such questions, based on ordinary scenarios, make it difficult to acquire the English skills needed for work-related tasks directly. For example, a programmer requiring English for API documentation gains little from fill-in-the-blank questions about family composition or reading comprehension exercises on threats to coral reefs in the Pacific Ocean.

%

%
To address the challenge of English studies being disconnected from workers’ everyday tasks, we draw on task-based language teaching~\cite{nunan2004task} and situated learning~\cite{lave1991situated}. When study materials are embedded in workers’ job contexts, they can improve task performance while sustaining engagement in language learning~\cite{anthony2018introducing}. Prior research shows that grounding instruction in authentic tasks not only supports second language acquisition~\cite{long1996role, Willis2021TaskBasedLearning}, but also enhances motivation~\cite{Hutchinson1987ESP} and strengthens memory retention~\cite{Coyle2010CLIL}. Building on these insights, we investigate whether generating study materials of mobile language learning apps directly from work tasks can further enhance language learning and foster long-term engagement.
\idx{1,2}\revised{Prior work has explored context-aware content generation using daily situations, objects, or personal interests, and demonstrated technological feasibility and learning benefits for general learners~\cite{edge2011micromandarin, hautasaari2019vocabura, draxler2023relevance, yamaoka2022experience}. 
However, little has been explored how to support workers in EFL learning, even though a large amount of personal data from information-work contexts remains underutilized despite its potential to generate personalized learning materials.}

\revised{To understand EFL workers’ current strategies and challenges across both their work and learning environments, we conducted} 
a formative study consisting of an online survey and follow-up interviews with 49 non-native information workers in South Korea.
\revised{
Participants expressed a strong willingness to improve their English proficiency for work, as lexical disruptions and low confidence in English frequently hindered their tasks.
As a result, they turned to easily accessible English learning mobile apps, but the disconnection of the study materials from their work contexts reduced perceived helpfulness and engagement.}
In addition, \revised{participants heavily relied on} LLM-based assistants, \revised{frequently asking to look up, translate, and proofread}. 
\revised{To improve work-related English proficiency, a small subset of participants engaged in self-directed review practices: They manually recorded unfamiliar vocabulary and revisiting unclear passages, often transferring these materials into flashcard or quiz apps.
However, participants reported that maintaining such manual practice routine was difficult, describing them as burdensome and taxing especially after work. }

\idx{2,7}
\begin{revisedenv}
Based on the lessons from the formative study, we aimed to reduce learner burden and support engagement in work-related EFL learning by streamlining review practices of work-related English use,
\cameraready{thereby shifting away from traditional EFL instructions---which often rely on decontextualized, curriculum-driven materials---toward a more contextualized, usage-driven pedagogical approach.} 
As a first step toward this goal, we leveraged workers' interactions with LLM-based assistants for language queries---questions that EFL workers ask for their text-based English tasks and rarely revisit afterwards---to provide an automated routine for practice.
We present \end{revisedenv} \sysname{}, an ensemble of intelligent systems designed to generate English quizzes from workers' LLM queries. \sysname{} consists of three components: (1) \queryapp{}, an \idx{17}\revised{LLM-based desktop} chatbot that answers workers' English-related queries~(\blackrectsmall{A} in \autoref{fig:teaser}); (2) \quizapp{}, a mobile app that allows workers to complete short, context-relevant quizzes at their convenience~(\blackrectsmall{C} in \autoref{fig:teaser}); and (3) the backend pipeline that processes queries from \queryapp{} to generate and validate quizzes~(\blackrectsmall{B} in \autoref{fig:teaser}). 
\revised{Leveraging conversation logs with \queryapp{} as a source material, the backend automatically produces a set of personalized multiple-choice questions for each user using an LLM. The questions then undergo AI-driven quality checking and refinement to ensure that each has a single correct answer and an appropriate level of difficulty. On \quizapp{}, users can solve quizzes consisting of 10 questions that mix newly generated and previously solved ones.}

We conducted a three-week field deployment with 28 information workers in South Korea. 
\idx{13}\revised{Through the deployment study, with a focus on feasibility, we explore the following research questions: \textbf{RQ1}--How do EFL workers engage in and sustain their English learning when using study materials generated from their work context? \textbf{RQ2}--How does studying English with work-related content influence workers’ learning outcomes and self-efficacy? and \textbf{RQ3}--How do EFL workers perceive the value of studying English with questions generated from work-related content?
}

\cameraready{Our results showed that participants consistently engaged with \sysname{} throughout the study period and perceived review practice with \sysname{} as more sustainable than their previous experiences with other study methods. 
Furthermore, our pipeline generated quality-assured questions that showed strong alignment with expert evaluations.
We observed that participants' self-efficacy in English skills increased significantly after actively using \sysname{} (9.5\% gain on average, $p<0.001$), and participants in the beginner-level English proficiency showed notable learning gain on a TOEIC-based English proficiency test.} 
In the post-study survey, participants reported that quizzes generated by \sysname{} were more relevant to their work and were helpful in their tasks compared to their prior experience of studying English. It also encouraged them to consider the educational aspect of their work-related queries. 

The key contributions of this work are as follows:
\begin{enumerate}[leftmargin=*, itemsep=4pt, topsep=2pt]
    \item \idx{3}\revised{Design and implementation of \sysname{}, an LLM-based system that supports lightweight review practice of daily English use by automatically generating quizzes from work-related English queries in an LLM-based assistant. \sysname{}'s design was informed by a formative study ($N=49$) with non-native information workers.} The source code of \sysname{} is publicly available at \sourcecode{}.
    \item \cameraready{Empirical findings from a three-week deployment study ($N=28$) showing that \sysname{} helps EFL workers sustain engagement and achieve measurable learning benefits through task-specific qualified quizzes with low burden, suggesting the potential of sourcing from on-the-fly LLM queries as an alternative to traditional decontextualized EFL learning methods.}
    \item \revised{Design considerations for AI-mediated EFL learning material generation that fosters engagement and proficiency by leveraging knowledge workers' growing reliance on LLMs.}
\end{enumerate}

%% file: sections/02_relatedwork.tex
\section{Related Work}
In this work, we cover the related work in the areas of (1) EFL learning for information workers, (2) creating context-aware learning materials, and (3) retrieval practice for second language acquisition.

\subsection{EFL Learning for Information Workers}
English plays a critical role in the workplace as the common language of global industries and disciplines~\cite{kirkpatrick2011internationalization,veerasamy2014teaching}.
The proliferation of digital work environments has posed unique challenges for non-native English-speaking workers, who navigate English-language texts as part of their computer-based tasks. For example, Amano \etal{} found that researchers spend 46.6\% more time reading English papers and 50.6\% more time writing them compared to native speakers, while facing higher rejection rates~\cite{amano2023manifold}. 
Similarly, non-native English programmers struggle with technical documentation, professional communication, and code comprehension~\cite{guo2018non}.
Consequently, English-as-a-foreign-language (EFL) learning has become critical for information workers to develop English proficiency---the ability to achieve goals through the use of English in relation to the specific purposes~\cite{Hutchinson1987ESP}.

In the field of HCI, previous research has explored approaches to directly support EFL workers’ use of English in the workplace, such as assisting with email composition, providing on-demand evaluations of writing, simplifying complex texts, and paraphrasing with AI explanations, in order to reduce context-switching burdens and disruptions during work~\cite{kim2025design,chen2024worddecipher, kwon2025better,dang2025corpusstudio,buschek2021impact,ito2023use,higasa2024keep}. 
However, effective support for EFL workers in the long run requires fostering English skills that directly enhance their ability to accomplish work tasks. In response, our work focuses on cultivating English proficiency within professional contexts.


Conventional English education, with its emphasis on general language proficiency, often falls short of meeting the specific needs of professional contexts. 
To address this gap, education theories such as task-based language teaching~\cite{nunan2004task} and situated learning~\cite{lave1991situated}  underscore the critical role of \textit{authentic} materials---texts and resources originated from real-world application~\cite{gilmore2007authentic}. 
Authentic materials give learners up-to-date domain information, unlike predefined materials that may lag behind current developments~\cite{benavent2011use}. Other studies found that working with content related to one’s professional field enhances motivation and self-efficacy by building confidence in handling real-world tasks~\cite{bielousova2017developing}. Context-specific, task-based materials help learners develop active knowledge directly applicable to their daily work~\cite{yamada2012development}.
Building on this emphasis, our work explores flexible ways to support EFL learning by providing authentic, work-related materials generated from workers’ daily English tasks.


\subsection{Creating Context-Aware Learning Materials}
Context is fundamental to learning because knowledge is inseparable from the situations and activities in which it is acquired~\cite{dewey2024democracy, collins2006cognitive}. Context-aware learning approaches~\cite{hwang2008criteria}, which situate learning in the learner’s personal context by selecting, adapting, or generating content, are particularly well-suited for language learning.
Because contextually relevant materials allow learners to experience language in authentic settings~\cite{lee2022systematic} and foster situational interests~\cite{hidi2006four}, they thereby enhance motivation and engagement~\cite{hidi2006four}, prevent inert knowledge~\cite{liu2009context}, and ultimately promote meaningful and effective learning~\cite{draxler2023relevance}.

The field of HCI has explored context-aware personalized learning materials that adapt to factors such as learners’ location~\cite{edge2011micromandarin,hautasaari2019vocabura}, surrounding elements~\cite{ibrahim2018arbis,draxler2023relevance}, social media content~\cite{coleman2012twasebook,yamaoka2022experience}, and other contextual information~\cite{ogata2014ubiquitous}.
For example, MicroMandarin~\cite{edge2011micromandarin} suggested flashcards relevant to nearby venues, while Vocabura~\cite{hautasaari2019vocabura} generated L1-L2 word pairs from walking commute routes, both leveraging GPS coordinates to support vocabulary learning. 
Draxler \etal{} further explored an object-based approach that automatically generates exercises by detecting elements in learner-captured photos from their daily contexts~\cite{draxler2023relevance}.
In addition, Yamaoka~\etal{} introduced a method that extracts keywords from Instagram posts to generate example sentences, helping learners acquire new words aligned with their interests and improving retention~\cite{yamaoka2022experience}.
Beyond academic research, recent language learning services have begun adopting generative AI to provide situated and contextualized learning experiences~\cite{googlelittlelessons,duolingomax}.




With the growth of digital environments, research has increasingly focused on usage-based learning to help learners acquire practical language skills by leveraging learner–computer interactions as sources of contextual content, such as eye gaze~\cite{ding2025unknown}, clicked hyperlinks~\cite{de2002visible,sanko2006effects}, translations~\cite{lungu2018we}, and other digital traces~\cite{azab2013nlp}.
Ding \etal{} identified unknown words through gaze trajectories while learners read foreign language texts, offering real-time translations and explanations for just-in-time vocabulary acquisition~\cite{ding2025unknown}.
Lungu \etal{} proposed a comprehension approach that generated mobile exercises from learners’ translated sentences during web reading, which could serve as potential learning cues~\cite{lungu2018we}.
This work demonstrated the feasibility of this ecosystem, highlighting engagement and learning benefits. 

Our work extends this line of research by leveraging emerging conversational interactions between learners and LLM-based chatbots. 
We argue that queries to LLMs reflect learners’ immediate language difficulties and learning intentions, serving as valuable cues for situated and usage-based learning.
In this work, we aim to generate work-related learning exercises from EFL workers’ queries collected in the course of their professional tasks.

\subsection{Retrieval Practice for Second Language Acquisition}
In second language acquisition, the \textit{practice} of difficult linguistic features offers learners opportunities for meaningful language use, reinforces task performance, and fosters adaptive language proficiency~\cite{thompson2019practice,johnson1997language}.
In particular, retrieval practice, which involves actively recalling knowledge from memory, typically through exercises such as quizzes or self-testing, is one of the most effective review strategies~\cite{rowland2014effect,toppino2018metacognitive}. 
It has been shown to be beneficial than simple `restudy' in strengthening long-term memory and supporting the transfer of knowledge to new context~\cite{rowland2014effect,rawson2011optimizing}.
Accordingly, prior work has integrated retrieval-based exercises into second language learning systems to enhance vocabulary acquisition, reading comprehension, and communicative fluency~\cite{culbertson2016social,draxler2022agenda,edge2012memreflex}.

When combined with microlearning~\cite{gassler2004integrated}---an approach that delivers educational content in small, easily digestible units---retrieval practice becomes particularly effective for busy adults.
Microlearning helps sustain motivation and engagement while minimizing the time burden, making it well-suited for learners balancing work and study~\cite{inie2021aiki,kim2024interrupting}. 
Many popular mobile-assisted language learning (MALL) systems, such as Duolingo~\cite{duolingo}, Anki~\cite{anki}, and Quizlet~\cite{quizlet}, leverage this principle by offering interactive exercises (\eg{}, flashcards, multiple-choice questions, and fill-in-the-blanks).
In addition, the field of HCI has explored retrieval-based microlearning in various contexts, such as spaced practice for vocabulary acquisition, adaptive exercise scheduling, and bite-sized practice integrated into daily routines~\cite{draxler2022agenda,edge2012memreflex}. 

Building on this, our work provides a mobile practice environment that leverages retrieval practice for information workers learning practical English skills.
We focus on a particular exercise type---multiple choice fill-in-the-blank questions---which are widely used in standardized proficiency tests~\cite{toeic}. 
Furthermore, recent research has demonstrated that AI-generated multiple-choice questions can reach expert-level quality~\cite{elkins2024teachers,doughty2024comparative}, highlighting their potential as a scalable and effective method for creating adaptive practice.
This allows us to focus on the content's quality, which we ensure through our systematic generation pipeline (see \autoref{sec:backend}).

%% file: sections/03_formative.tex
\section{Formative Study}
To inform the design of \sysname{}, we conducted an online survey with 49 information workers whose native language is Korean, followed by semi-structured interviews with ten volunteers. 
We aimed to understand the type of barriers they face during daily English-related tasks, limitations with the existing digital tools they use to handle these tasks, and effective EFL learning practices they have experience using to develop work-related English skills.


%

\subsection{Procedure and Analysis}
\ipstart{Online Surveys}
Through both closed and open questions, we asked participants about the challenges they face as non-native English speakers at work, the digital tools they use to support English-related tasks, the effectiveness of their EFL learning strategies, and their perceived need for continued learning. 
We also asked about the willingness to participate in a follow-up interview.
The online survey was advertised to native Korean speakers on social media and our internal network, inviting information workers who use computers for their work and regularly perform tasks that require English. 
Forty-nine people (25 females; aged 22--49) completed the survey, which included 22 researchers, 11 engineers, and 16 professionals from various fields, including strategic planning, sales and marketing, design, general affairs, and healthcare. 
The survey took approximately 20 minutes to complete.
We compensated 5,000 KRW (approx. 4 USD) for survey respondents.

\ipstart{Interviews}
For in-depth analysis, we conducted follow-up interviews with ten survey respondents who indicated their willingness to attend as part of the online survey. 
Each interview lasted about 40 minutes and was conducted in person or remotely, depending on the participants' availability. 
We revisited the interviewees' survey responses and asked them to elaborate on their open-ended answers.
Using screen sharing and think-aloud protocols~\cite{cooke2010assessing}, participants walked through recent scenarios involving English-related tasks, demonstrating queries they had made to generative AI (\eg, ChatGPT, Gemini) or other tools. They also described their review practices focused on work-specific English content. We compensated 20,000 KRW (approx. 14 USD) for interview participants.

\ipstart{Analysis}
All interviews were audio-recorded and transcribed for analysis.
We summarized the closed-ended survey questions using descriptive statistics. 
We used Thematic Analysis~\cite{braun2006using} to qualitatively analyze both the open-ended questions of the survey and interview transcripts. One researcher coded survey responses as well as interview transcripts simultaneously, grouping them into broader themes.
The research team iterated through several rounds of discussion to refine these themes.
In the following sections, we present findings from both the survey and the interviews, referring to each interview participant as I1 through I10. 

\subsection{Finding 1: Understanding the Difficulties of EFL Workers in the Workplace}
Participants worked with large amounts of information written in English as part of their daily tasks, ranging from (1) communication through emails or messengers, (2) accessing online resources, and (3) writing professional documents such as reports or papers, \idx{7}\revised{most of which required intensive reading and writing rather than spoken communication.} 

One common linguistic difficulty that the majority of respondents (25/49) pointed out was \textbf{lexical disruption}, noting that unfamiliar domain-specific terminology often hindered their comprehension and prompted them to look up words frequently. 
I5, who works in governance administration at an international research lab, remarked that ``\textit{The official materials from the UN Headquarters are often overly formal and full of UN-specific terms, which slows me down as I have to look them up.}'' 
Similarly, I1---an international business development manager---noted ,``\textit{For example, I used to think  `airway' only meant a flight route, but later learned it also refers to a respiratory tract [a human body part].}''
Relatedly, participants reported not only linguistic challenges but also affective challenges---stemming from a lack of confidence, which consequently hindered their workflow. 
Four participants 
mentioned in their surveys and interviews that they proofread emails for grammar, formality, and tone before sending, concerned that mistakes might appear impolite or give a negative impression of their professional competence.

Participants often felt that current EFL learning was \textbf{disconnected from work context}. 
More than 2/3 of the survey respondents (33/49) rated that they often (45.0\%) or always (22.5\%) feel the need for learning English for work on a 5-point Likert-type scale question.
Yet, the majority of them (28/49) were not currently studying English, demonstrating the difficulty of constant engagement in EFL learning practices during work. 
For those who were currently studying English, all of them (21/21) reported that they used easily accessible and self-directed mobile apps, outside of work context (\eg{}, Duolingo~\cite{duolingo}, Speak app~\cite{speakapp}). Other common practices included online tutoring (\eg{}, Ringle~\cite{ringle}; 8/21), reading English novels or articles (8/21), shadowing (4/21), and in-person courses (3/21). 
However, the majority of them (14/21) struggled to sustain their learning because irrelevant learning materials did not translate into practical support in their work contexts. In the follow-up interview, I10 noted, ``\textit{What I really want to learn right now is material I can use immediately in business meetings, but finding a suitable platform or tool has been very difficult}.'' I1 also remarked, ``\textit{I'm often exposed to highly specialized medical terms, but when the material is from a learning app or an article outside my field, it tends to use more general vocabulary and expressions. While this is helpful for conversations, it's not very useful when reading work-related articles or clinical papers.}''

Challenges occurred during the reviewing phase as well due to \textbf{lack of sustainable review routines}. 
To align EFL learning with their work contexts, six interviewees once tried to review unfamiliar words and expressions from work by compiling personal glossaries and organizing them with tools such as Notion, Google Docs, or the open-source flashcard app Anki~\cite{anki}.
However, participants failed to maintain engagement with such review routines, as manually collecting work-related vocabulary or expressions was time-consuming and burdensome.
Moreover, reviewing these materials with explicit exercises further discouraged continued practice. 
In the follow-up interviews, I9 noted, ``\textit{After work, I don't want to revisit the traces of what I did during the day. Reviewing would mean opening my daily logs in a workspace like Notion, finding the target words, gathering them on another page, and then asking GPT or searching Google for their meanings. Most days, it just feels too much.}''
Also, I10 remarked, ``\textit{In one-on-one business English tutoring, my teacher listed my mistakes in Google Docs, but reviewing them felt like just reading meeting minutes and was neither fun nor motivating. I wish I could review them in more engaging ways, like quizzes or other formats for sustainable practice.}''
While interactive apps such as Anki, with flashcard and quiz features, were available, participants (I8, I9) found it overly complex and overwhelming to customize and manually upload word lists, especially after work.

\subsection{Finding 2: Common Patterns of English Language Queries}
To address language barriers, all participants used language assistance tools for lookup and double-checking, including dictionaries, web search engines, translators, AI-based writing assistants (\eg{}, Grammarly~\cite{grammarly}, DeepL~\cite{deepl}), and LLM-based chatbots (\eg{}, ChatGPT~\cite{chatgpt}, Gemini~\cite{gemini}, Claude~\cite{claude}). 
In particular, most survey respondents (46/49) commonly used LLM-based chatbots, which offered convenient conversational support and context-aware explanations for a wide range of English-related difficulties.
Interviewees entered queries primarily by copying and pasting text, ranging from single words to full passages, along with a recurring prompt for linguistic support. 
From their usage scenarios, we identified three prominent query patterns to LLM-based chatbots: \textbf{look-up}, \textbf{translation}, and \textbf{proofreading}.

Most interviewees (7/10) often used chatbots just like dictionaries to \textbf{look up} definitions of unfamiliar words or description of confusing grammar during their tasks.
I5 noted, ``\textit{These days I just ask ChatGPT when I'm unsure about grammar, like  `an MBA or a MBA?'}''
I6 and I8 found LLMs useful for clarifying domain-specific terms or subtle nuances, as they offered context-specific explanations, especially for words with multiple possible meanings. 

All interviewees (10/10) used chatbots to \textbf{translate} text in English to Korean and vice versa.
They translated text in English into Korean to ensure their comprehension and translated Korean into English to compose professional writing more efficiently.
I6 noted, ``\textit{I usually ask LLMs to align the original and the translation side by side, so I can double-check whether each part conveys the intended meaning,}'' highlighting the need for a dual-language view to enable rapid comparison under time pressure at work.

The majority of interviewees (6/10) also frequently asked chatbots to \textbf{proofread} their own draft, ranging from formal business documents to casual conversation with colleagues. 
Participants refined grammar, tone, and style to fit the context of the communication, often by providing additional information (e.g., their relationship with the interlocutor) to the assistant.
For example, I5 copied entire email threads into the LLMs and asked, ``\textit{Please proofread my reply,}'' to ensure their response was both grammatically sound and aligned with the ongoing exchange.
I9 even checked short messages for online meetings, such as ``Will you be joining soon?'', to understand the nuance of the message that they may be implying in the message: ``\textit{I worry it might sound like I'm pushing, so I ask ChatGPT to review even simple texts before sending.}''

%% file: sections/04_system.tex
\section{\sysname}
Our formative study revealed that EFL workers suffer from \textit{lexical disruption} while handling information work in English. In addition, our participants noted that most existing EFL learning systems rely on generic materials disconnected from their work contexts---despite research in EFL education showing that authentic, usage-based practice fosters learner engagement and improves both proficiency and self-efficacy~\cite{gilmore2007authentic,benavent2011use,bielousova2017developing}.
To address these challenges, we designed and developed \sysname{}, a language querying and self-directed learning system that provides work-related quizzes generated from language queries. In this section, we discuss our design rationales from the formative study and literature. We then describe our system design and generative pipelines, along with implementation details.

\subsection{Design Rationales}
\ipstart{DR1. Leverage AI-Assisted Language Queries as a Source of Learning Material} 
In our formative study, nearly all participants relied on LLM-based AI assistants for work-related English tasks, such as looking up unfamiliar terminology, resolving confusing grammar, translating text, or proofreading their writing.
We therefore treated users' language queries with an AI chatbot as an authentic source from which we can learn the English assistance that they need and generate learning materials. 
Moreover, certain types of user queries, such as searching for a definition of a word or comparing input text with edited text, explicitly reveal their weakness in English proficiency.

\ipstart{DR2. Optimize AI Assistant Interface for Language Querying} 
Since our formative study participants frequently used generative AI assistants, such as ChatGPT, for language practice, we observed that their querying interactions were often inefficient and tedious. For example, participants had to repeatedly type boilerplate commands in the input (\eg{}, ``\texttt{Translate this into Korean:}'') whenever they initiated a new query. In addition, participants often issued follow-up requests to format responses for their language tasks (e.g., requesting to display Korean and English text side by side or highlighting edited portions to track changes), which led to unnecessary back-and-forth dialogue turns.

\begin{figure*}[t]
    \centering
    \includegraphics[width=\textwidth]{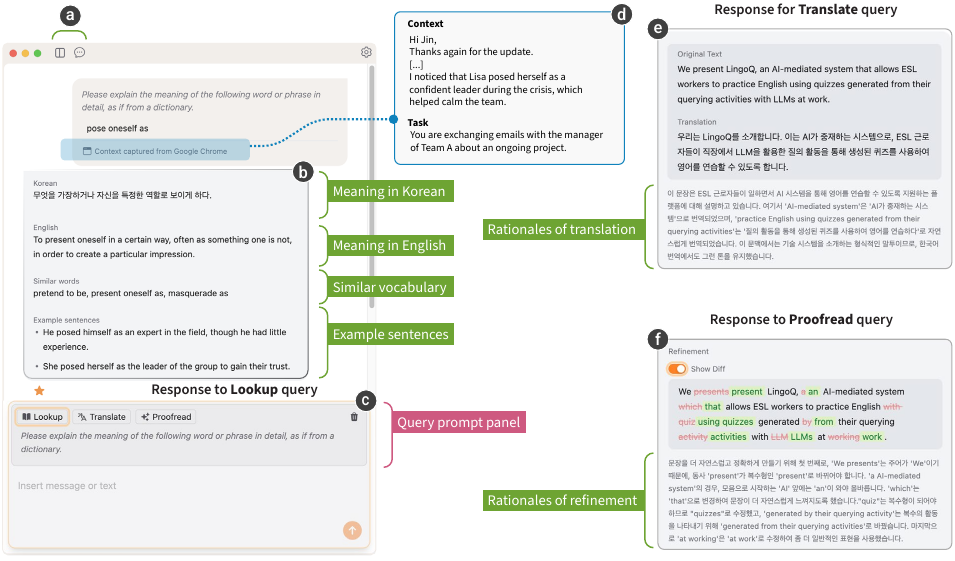}
    \caption{Main window and the interface components of \queryapp{}. Users can open new chat threads of their choice in the thread list or via the New Chat button (\circledigit{a}). The AI's responses for three major types of query intents---\textit{Look up} (\circledigit{b}), \textit{Translate} (\circledigit{e}), and \textit{Proofread} (\circledigit{f})---provide the UI components tailored to each query type. The new message panel (\circledigit{c}) incorporates a query prompt panel at the top. Users can insert a template query prompt in their message via the three quick-access buttons. When a query is sent via a keyboard shortcut\shortcut{}, the system analyzes a screenshot of the user's active window, and the user can review both the surrounding context of the copied sentence and the task at the time of the query \circledigit{d}. By clicking the star icon \starmark below AI responses (\circledigit{b}), users can mark the message to increase the likelihood that the corresponding question will be included in \quizapp{}. 
    }
    \Description{Actual screenshots of LingoQuery. (a) Two buttons on the top of the UI window, with a sidebar expand icon and a text balloon icon for opening a new thread. (b) An example response from \quizapp{} displayed in a Dictionary format container. Korean expression, English expression, Synonym, and Example Sentences are displayed from top to bottom. (c) A chatbox that a user can use to enter a query. Three buttons with labels Lookup, Translate, and Proofread are above a text input box. (d) An annotation that contains an email draft, to represent the context from which a user looks up a word. (e) An example response from \quizapp{} displayed in a Translation format container. On top, the translated text in English is present, and the original text in Korean is displayed right below. Under the Korean text, the rationale behind the translation is present in a separate container. (f) An example response from \quizapp{} displayed in a Proofread format container. On top, a toggle button that can turn on track change. The button is on, and the proofread text has deleted words in red with strikethrough and added words in green. Under the container that has the proofread text, the rationale behind the editing is present in a separate container. }
    \label{fig:interface:lingoquery}
\end{figure*}

Hence, we incorporated \queryapp{}, an LLM-based assistant dedicated to language queries. \idx{1}\revised{Leveraging the design of typical AI assistant chatbots like ChatGPT~\cite{chatgpt} or Gemini~\cite{gemini}, which information workers are already familiar with, we adopted a similar structure while optimizing the interactions and interfaces for such usage.} Given that participants in our formative study often copied text to query digital tools, we implemented a keyboard shortcut that directly copies and pastes selected text from the computer into a new chat message. We also introduced \textit{query intents} that users can attach to an input message, which automatically insert predefined yet customizable prompts for three frequent request types---look up, translate, and proofread---thereby avoiding manual typing of boilerplate instructions. The AI responses to these query intents are rendered in language-relevant message displays (see~\autoref{fig:interface:lingoquery}).


\ipstart{DR3. Streamline Reviewing Work-Related Language Activity} 
Participants in our formative study attempted to review vocabulary or expressions used in daily tasks at work, but the burden of manually collecting and revisiting materials without explicit exercises hindered sustained engagement, particularly after work.
Inspired by literature suggesting that microlearning with short practice sessions embedded into daily routines can foster sustained engagement and improve proficiency~\cite{kim2024interrupting,inie2021aiki,gassler2004integrated}, we designed our system to generate bite-sized interactive quizzes. 
These quizzes are generated directly from automatically collected queries, supporting continuous practice without extra burden outside work.
We adapted multiple-choice fill-in-the-blank question formats from standardized tests (\eg{}, TOEFL, TOEIC, GRE) to support vocabulary and grammar practice. 
While the format of problems can be diversified (\eg{}, reading/listening comprehension, open-ended questions, short writing tasks), we limit our study material to the fill-in-the-blank multiple-choice question (MCQ) format for its simplicity to support easy access to English study anytime, anywhere on a mobile phone. The effectiveness of other question formats lies beyond the focus of this work, which is to generate study material relevant to work. 

Based on these design rationales, we developed \sysname{} that consists of \queryapp{} (\ref{sec:query}), \quizapp{} (\ref{sec:quiz}), and the backend pipeline (\ref{sec:backend}). \queryapp{} is a desktop-based AI assistant that users can share English-related queries or discuss freely.  The backend pipeline manages user query data from \queryapp{} and generates questions and curates them into quizzes. \quizapp{} is a mobile application that offers 10-question quizzes generated from the user's dialogues with \queryapp{}.

\subsection{\queryapp{}}
\label{sec:query}
\subsubsection{Interaction Components of \queryapp{}}

\queryapp{} adopts the typical interface design of desktop versions of LLM-based AI assistants, such as ChatGPT~\cite{chatgpt} and Claude~\cite{claude}, while incorporating bespoke interaction components tailored to the English-language query contexts. 
The sequence of chat messages is organized into chat threads, and users can either start a new thread or append messages to existing ones by selecting them from the sidebar (\circledigit{a} in~\autoref{fig:interface:lingoquery}).
By default, the AI response messages are rendered as a markdown-formatted view.

\ipstart{Language Query Intent Selection and Prediction}
The system supports three predefined query intents: (1) \textit{Look up}, (2) \textit{Translate}, and (3) \textit{Proofread}.
When composing a new message, users can explicitly select a query intent in the chatbox by pressing the buttons at the top, which load a predefined prompt (\circledigit{c} in~\autoref{fig:interface:lingoquery}). 
Each query intent applies a prompt template that is concatenated with the user's message input; for example, ``\texttt{Please explain the meaning of the following word (or expression) in detail in dictionary format}'' for Look up, which the user can edit before sending.
If no query intent is selected, the message in the chatbox is treated as a plain prompt in text, and the system automatically predicts its query intent when generating a response. 
If a user's query corresponds to one of the three query intents, the system offers a customized view that highlights the structured information of the intent type. 
The \textit{Look up} response type follows a typical dictionary format (\circledigit{b} in \autoref{fig:interface:lingoquery}); the \textit{Translate} response type provides a side-by-side view for comparing original and translated text (\circledigit{e} in \autoref{fig:interface:lingoquery}); and the Proofread response type displays a formatted container showing the proofread text with the rationale for edits underneath, along with an option to toggle track changes so users can quickly see where edits were made (\circledigit{f} in \autoref{fig:interface:lingoquery}). These views were informed by how interviewees in the formative study customized responses when using generic LLM-based assistants.

\ipstart{Shortcuts and Contexts}
To enable users to receive assistance within the context of work-related applications, \queryapp{} provides an operating system–level shortcut to trigger a query. When the user highlights text anywhere on the computer and presses `\texttt{Ctrl} + \texttt{Cmd} + \texttt{C}' on MacOS or \texttt{Ctrl} + \texttt{Alt} + \texttt{C} on Windows, the \queryapp{} window opens with a new chat thread and the copied text pre-filled in the chatbox. At the same time, the system captures a screenshot of the active window and displays it alongside the text when the shortcut is pressed. Before submitting the query, the user can choose to include the screenshot with the message or remove it if it contains sensitive information. If the screenshot is included, the system runs image understanding on OpenAI's \texttt{GPT-4o} model to extract the text surrounding the copied content and infer the nature of the tasks based on metadata of the visible application on a screen, enriching the content for question generation later. The inferred context will be displayed in the UI as well.

\ipstart{Marking Messages for Prioritizing Question Generation}
Users can also mark noteworthy AI responses via the \starmark star icon (right above \circledigit{c} in~\autoref{fig:interface:lingoquery}). Marked messages are processed in the system pipelines, increasing the likelihood that the corresponding question will be included in a quiz. This mark feature allows users to flag particular words or expressions they wish to review within the context of \queryapp{}, removing the need to manually track what they want to study.

\begin{figure*}[t]
    \centering
    \includegraphics[width=\textwidth]{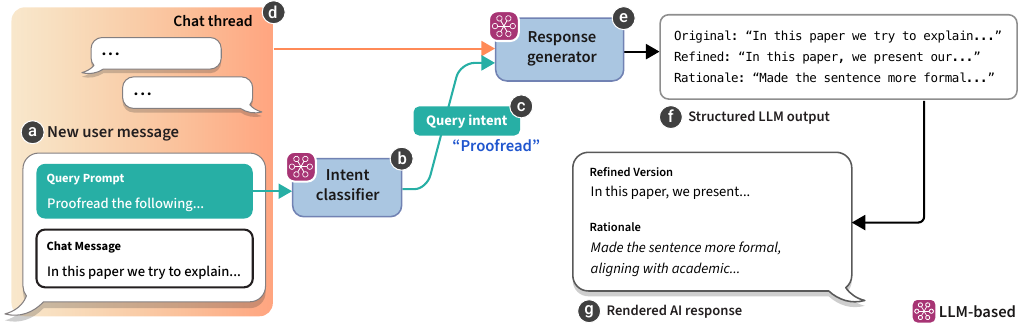}
    \caption{Conversational pipeline of \queryapp{}. When the user sends a new message \circledigit{a}, the intent classifier~\circledigit{b} identifies the query intent~\circledigit{c}, which is then passed to the response generator~\circledigit{e} together with the chat history~\circledigit{d}. The response generator produces an appropriate response~\circledigit{f} structured according to the query intent. Finally, \queryapp{} renders this structured response accordingly~\circledigit{g}.}
    \Description{A flow diagram illustrating the conversational pipeline of LingoQuery. The process begins with a user message, which is sent to an intent classifier that detects the type of query. The identified intent and the chat history are passed to a response generator, which produces a structured reply based on the intent. Finally, LingoQuery renders the structured response back to the user.}

    \label{fig:pipelines:conversation}
\end{figure*}

\subsubsection{\queryapp{} Conversational Pipelines}
\queryapp{} is a self-contained app that generates responses to user requests, similar to an LLM-based AI assistant with additional customization for specific query types. A user's query input is processed before being sent to the LLM engine (OpenAI API in our case). \autoref{fig:pipelines:conversation} illustrates the response generation pipeline of the \queryapp{} conversational agent when it receives a new user message (\circledigit{a} in \autoref{fig:pipelines:conversation}). 
LLM-based \textbf{Intent Classifier} (\circledigit{b} in \autoref{fig:pipelines:conversation}; see \autoref{appendix:prompts:query:intent} for the instruction provided to the LLM) determines the corresponding query intent (\circledigit{c} in \autoref{fig:pipelines:conversation}). Both the chat history (\circledigit{d} in \autoref{fig:pipelines:conversation}) and the detected intent are then passed to the \textbf{Response Generator} (\circledigit{e} in \autoref{fig:pipelines:conversation}), which produces an AI response using an LLM (See \autoref{appendix:prompts:query:response} for the instruction provided to the LLM).
When the query intent does not correspond to a plain-text message but instead falls into one of the intents Look up, Translate, or Proofread, the LLM output is returned as a JSON object containing relevant attributes (\eg{}, the \textit{original} input, \textit{refined} text, and the \textit{rationale} of refinement in the case of a Proofread intent). The structured format enables the application interface to render the output through a bespoke UI (\circledigit{g} in \autoref{fig:pipelines:conversation}).

\subsection{Question Generation Pipelines}
\label{sec:backend}

The question generation in \sysname{} follows three pipelines---question generation, question quality evaluation for filtering, and question selection---that transform interactions collected through \queryapp{} into validated quiz questions for \quizapp{}.

\subsubsection{Generating Questions from Conversation and Context}\label{sec:pipeline:question}
\begin{figure*}[t]
    \centering
    \includegraphics[width=\textwidth]{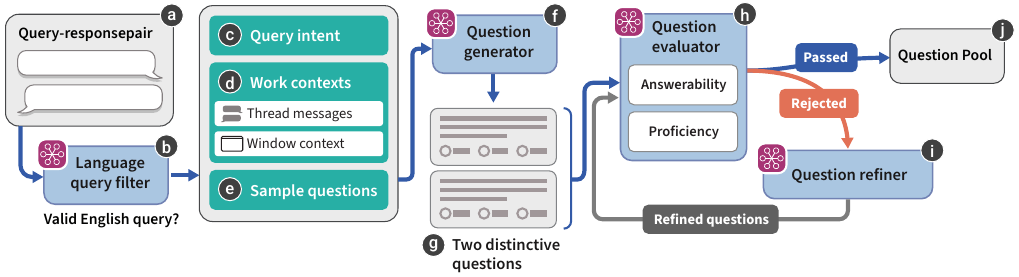}
    \caption{Question generation pipeline of \sysname{}. When a query–response pair arrives \circledigit{a}, the language query filter \circledigit{b} identifies the query intent \circledigit{c}, which is then passed to the \textbf{Question generator} \circledigit{f} together with work contexts \circledigit{d} and exam samples \circledigit{e}. The generator produces two candidate questions \circledigit{g}, which are evaluated by the \textbf{Question evaluator} \circledigit{h} on two criteria: answerability and proficiency. The \textbf{Question refiner} \circledigit{i} refines each question up to two iterations, and items that still fail are discarded. Accepted questions are stored in the question pool \circledigit{j}.}
    \Description{A flow diagram showing the question generation pipeline of LingoQ. (a) Query–response pairs enter the system and pass through (b) a language query filter that identifies (c) the query intent. The question generator (f) combines this with (d) work contexts and (e) exam samples to create two candidate questions (g). These are evaluated by (h) the \textbf{question evaluator} on answerability and proficiency. The \textbf{question refiner} (i) refines questions up to two times; items still failing are discarded. Accepted questions are stored in (j) the question pool.}
    \label{fig:pipelines:generation}
\end{figure*}

\autoref{fig:pipelines:generation} illustrates the question generation pipeline, which periodically produces fill-in-the-blank multiple-choice questions from query-response pairs collected in \queryapp{} and captured context data.
Every five minutes, the system checks for new query–response pairs collected from \queryapp{} (\circledigit{a} in \autoref{fig:pipelines:generation}). For each pair, an LLM-based module evaluates whether the query is English-related (\circledigit{b} in \autoref{fig:pipelines:generation}); if not, it is filtered out from the question generation pipeline.
For eligible queries, the system takes the eligible pair and additional information---the conversation history with contextual data captured from the screenshot---to initiate generation (\circledigit{d} in \autoref{fig:pipelines:generation}).
The pipeline applies a different system prompt with few-shot examples modeled after TOEFL, TOEIC, and GRE example questions (\circledigit{e} in \autoref{fig:pipelines:generation}). 
For each conversation, two distinct questions are generated to ensure variety (\circledigit{f} in \autoref{fig:pipelines:generation}). 
Each question will generate one structured output in JSON format that contains: stem with a blank, key, distractors, explanation, and rationale for question generation (\circledigit{g} in \autoref{fig:pipelines:generation}). 

Finally, the contextual information extracted from a screenshot and conversation history is fed into the question generation pipeline to produce questions aligned with that context. The question generation module ensures that the question stem is relevant to the worker's task context. For example, a user query may simply involve searching for a word (e.g., ``airway''), in which case the surrounding text (e.g., ``the patient’s airway to ensure proper breathing'') from the screenshot can be used to extract the context and generate a relevant question stem. When a user submits an answer in \quizapp{}, the explanation provided will include this context to supplement the rationale for the correct answer.



\subsubsection{Quality Assurance of Generated Questions}
\label{sec:quality}
To ensure the quality of generated questions, they are evaluated by an LLM-based evaluation module informed by prior literature~\cite{elkins2024teachers,doughty2024comparative}. The evaluation applies two binary criteria: answerability, that is, whether the question can be clearly and correctly answered, and proficiency, that is, whether the question requires an appropriate level of English skill and is not too easy to answer (\circledigit{h} in \autoref{fig:pipelines:generation}). Questions that fail one of the criteria are iteratively refined—up to three times in total—by feeding the evaluator's rationale for failure, along with the original input, back into the question generation module (\circledigit{i} in \autoref{fig:pipelines:generation}). If a question passes within three iterations, it is added to the question pool (\circledigit{j} in \autoref{fig:pipelines:generation}), otherwise discarded. 

\subsubsection{Question Pool and Selection Logic}
After quality checks, questions undergo a final format validation to ensure that all required components—stem with blank, distractors, key, and explanation—are properly structured. 
Validated questions are then added to the question pool.
When a user initiates a quiz, ten questions are drawn from the pool. 
Each quiz contains 10 questions: 7 are selected from newly generated questions, and the remaining 3 are randomly drawn from the pool of questions that a user has solved previously using weighted probability. The system assigns higher weights to questions that have been repeated less frequently, answered incorrectly in the past, marked with the \starmark star icon in \queryapp{}, or not practiced recently. As a result, each quiz balances new questions (70\%) with questions that workers need to revisit.
The examples of generated questions for each type are available in \autoref{fig:question_examples}. 

\begin{figure*}[t]
    \centering
    \includegraphics[width=\textwidth]{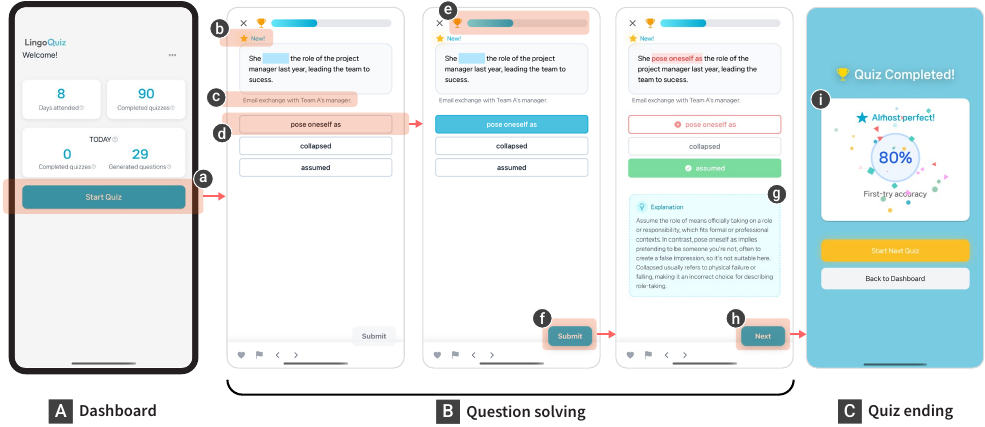}
    \caption{Main screens of \quizapp{}. In the Dashboard screen \blackrectsmall{A}, users can check their records and stats, along with the number of new questions add to their question pool today. When starting a quiz by pressing the Start Quiz button \circledigit{a}, a quiz with 10 unique questions are provided sequentially \blackrectsmall{B}.
    The new question that appears to the user for the first time is indicated by the star icon (\circledigit{b}). For questions generated from messages with context, the task description is provided (\circledigit{c}). To solve the question, the user can select an option (\circledigit{d}) and press the Submit button (\circledigit{f}) to submit an answer. Then the result the question is shown immediately, with explanation (\circledigit{g}), regardless of whether the user had selected a correct answer or not. After solving the ten questions, questions with wrong answers appears again, until all are answered correctly. The progress bar (\circledigit{e}) indicates the current progress. In the Ending screen \blackrectsmall{C}, users can practice a new quiz or return to the Dashboard screen.}
    \Description{Actual screenshots of LingoQuiz showing three main screens. (A) Dashboard with usage records. (B) Quiz interface where users solve ten questions until correct, with marked or new items, contextual reminders of the task, explanations shown after selecting an option, and a progress bar tracking completion. (C) Completion screen where users can start a new quiz.}
    \label{fig:interface:lingoquiz}
\end{figure*}

\subsection{\quizapp{}}
\label{sec:quiz}

Users can practice work-related English vocabulary and grammar by solving questions in \quizapp{}, generated from their dialogues in \queryapp{}. 
\quizapp{} provides a dashboard (\blackrectsmall{A} in \autoref{fig:interface:lingoquiz}) that helps users track how many quizzes they have completed that day, how many they have completed in total since the beginning, and how many new questions are available in the question pool (\circledigit{1} in \autoref{fig:interface:lingoquiz}).
Clicking `Start Quiz' launches 10 multiple-choice questions, each requiring users to fill a blank (\blackrectsmall{B} in \autoref{fig:interface:lingoquiz}).
When the question is generated from a marked AI response or when it is the first attempt, it is displayed with a \starmark or `new!' badge (\circledigit{2} in \autoref{fig:interface:lingoquiz}).
Questions generated from screenshots or sufficient thread context include the inferred task as a hint (\circledigit{3} in \autoref{fig:interface:lingoquiz}).
When users press the Submit button after selecting an option, the app provides immediate feedback on whether the answer is correct and the explanation (\circledigit{4} in \autoref{fig:interface:lingoquiz}).
Within the quiz, any incorrect questions reappear until users provide the correct answer.
Once all ten questions are answered correctly, the progress bar completes (\circledigit{5} in \autoref{fig:interface:lingoquiz}), and the quiz ends with a completion screen (\blackrectsmall{C} in \autoref{fig:interface:lingoquiz}).
Afterward, users may proceed to a new quiz or return to the dashboard (\blackrectsmall{A} in \autoref{fig:interface:lingoquiz}).


\subsection{Implementation}
We implemented the core system in Python running on a FastAPI~\cite{FastAPI} server that provides REST APIs for both \quizapp{} and \queryapp{}.
The chat history, generated quizzes, and user interaction data are stored in a PostgreSQL~\cite{PostgreSQL} database on the server.
The conversation and the question generation pipelines leverage OpenAI's Chat Completion APIs~\cite{openai} on top of the LangChain~\cite{LangChain} framework to run the underlying LLM inferences. All LLM inference and image understanding tasks are performed using a \texttt{gpt-4o} model. To protect user queries that may contain sensitive information, we used OpenAI Enterprise, which neither uses our data for training nor retains them.

We built \queryapp{} as a cross-platform desktop application using Electron~\cite{electron}, to support both Windows and MacOS desktop computers. The \quizapp{} app was implemented using React Native~\cite{ReactNative} as a cross-platform mobile application running on both iOS and Android phones. Both apps were written in TypeScript~\cite{TypeScript} and communicate with the server via REST API.

%% file: sections/05_userstudy.tex
\section{Deployment Study}
\input{tables/table-participants}
We conducted a three-week field deployment study with 28 EFL workers.
\revised{To address our research questions,} we aimed to examine how EFL workers engage with \sysname{} and how it affects their English proficiency and self-efficacy.
In addition, we conducted an expert evaluation to assess the quality of questions generated by \sysname{}. 
The study was conducted in South Korea with Korean native speakers and approved by the Institutional Review Board.

\subsection{Participants}\label{sec:participants}

\revised{We conducted power analysis to calculate the number of participants necessary. With an expected medium effect size of 0.5, the required sample sizes are 27 for a paired t-test (α = .05, one-tailed, power = .80) and 28 for a Mann–Whitney test (α = .05, one-tailed, power = .80).}
Our inclusion criteria were information workers who are: (1) working at least 30 hours per week, (2) using a computer as the primary work tool, (3) regularly performing tasks involving English, such as information access, communication, and document writing, (4) being a native Korean speaker, and (5) being an EFL learner.
We advertised our study on social media.
\revised{Initially, we recruited 34 workers to account for potential attrition; two participants dropped out due to their corporate security policies that prevented them from installing \sysname{}, and four were later excluded during analysis for not meeting the minimum requirements.}
Finally, a total of 28 EFL workers (\autoref{tab:participants}; P1–P28; 18 females and 10 males) completed the study.

Participants were aged between 25 and 48 years old ($M=33.4$, $SD=6.5$) and represented diverse professional domains.
Based on CEFR~\cite{council2001common}\footnote{The Common European Framework of Reference for Languages (CEFR) defines proficiency levels as basic (A1: beginner; A2: elementary), independent (B1: intermediate; B2: upper-intermediate), and proficient (C1: advanced; C2: proficient).} self-assessed English proficieny, 3 participants identified themselves as \textit{A1} (beginner), 4 as \textit{A2} (elementary), 14 as \textit{B1} (intermediate), and 7 as \textit{C1} (advanced).
All participants reported using LLM-based chatbots daily during their workdays.
Additionally, they had prior experience with EFL learning for work, including tutoring (17), vocabulary apps (13), and English media (12).
As a minimum requirement for study completion, we instructed participants to use \queryapp{} for at least 10 days during the three-week period and \quizapp{} for at least 10 days during the same period. To qualify as having used an app on a given day, participants needed to submit at least two questions in \queryapp{} and complete at least one quiz in \quizapp{}.
As compensation for their participation, we offered 200,000 KRW (approx. 144 USD) based on the required system usage over three weeks.

\subsection{Procedure}

\ipstart{Pre-Study Preparation}
Upon sign-up, we sent participants a link to a pre-study survey and a pre-study English proficiency test.
The survey included three items on a 5-point Likert scale that assessed the perceived relevance, effectiveness, and engagement of their past EFL learning methods, along with 16 items from the Questionnaire of English Self-Efficacy (QESE; eight on reading and eight on writing) on a 7-point scale, excluding speaking and listening to align with our research focus~\cite{wang2014psychometric}.

The English proficiency test consisted of 28 multiple-choice items selected from TOEIC (Test of English for International Communication)~\cite{toeic}, a standardized English proficiency test for general business.
\idx{7}\revised{We did not include spoken English proficiency measures because they lie beyond the focus of this work.}
The test included two types of questions: 16 simple fill-in-the-blank items---each with a single sentence and one blank---and three sets of four fill-in-the-blank items, each requiring participants to complete blanks within a single paragraph.
Before the study, we finalized the items from 46 questions by administering them to 29 information workers---who are not our study participants---and selecting those whose percentage of correct answers fell between 40\% and 80\%, as suggested in Classical Test Theory~\cite{devellis2006classical}. (see \autoref{appendix:test_validation} for details of the item validation.)

\begin{figure*}[t]
    \centering
    \includegraphics[width=\textwidth]{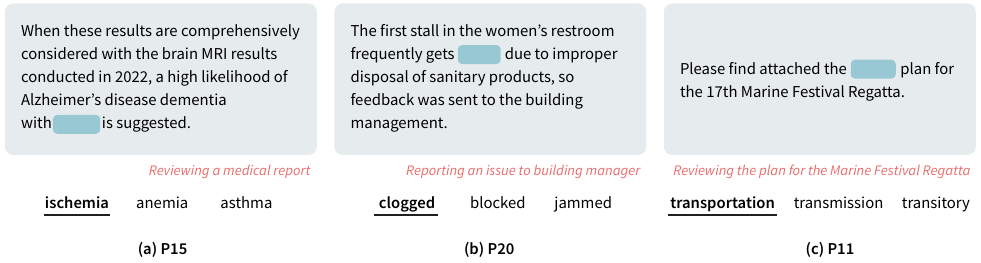}
    \caption{Selected questions actually generated by \sysname{} during the deployment study for three participants. Each question consists of a stem with a blank, context reminding the task at the time of query (red text), and three alternatives with correct answers underlined.}
    \Description{Three example multiple-choice questions generated from LingoQ during the deployment study. Each question includes a stem with a blank to be filled, contextual information reminding the user of the task at the time of query, and three alternatives with the correct option marked. The examples come from participants P15, P20, and P11.}
    \label{fig:question_examples}
    \labelphantom{fig:question_examples:p15}
    \labelphantom{fig:question_examples:p20}
    \labelphantom{fig:question_examples:p11}
    
\end{figure*}

\ipstart{Onboarding Session}
A group of 2-3 participants attended a 1-hour, in-person onboarding session, bringing their laptop to install \queryapp{}.
After explaining our study goals, we assisted participants with installing the system on their laptops and mobile devices. 
To ensure participants fully understood how to use the system, we conducted a hands-on tutorial that covered the main features of \sysname{}. 
This step ensured that participants had quizzes available during the early stage of the deployment period.

\ipstart{Deployment}
Immediately after the onboarding session, participants began using \sysname{} for three weeks. 
During this period, participants were instructed to direct their English-related queries to \queryapp{}, instead of using ChatGPT or other tools.
They were instructed to use \quizapp{} at any time of the day.

At the end of each week, we sent participants a message summarizing \sysname{} usage to remind them of the minimum requirement for study completion.
Additionally, when participants were inactive for more than three days, \quizapp{} sent an evening push notification. 
\quizapp{} notified participants if new questions were available for a day. 

\ipstart{Post-study Survey}
After the 3-week deployment period, we sent an online post-study survey and a post-study English proficiency test.
The survey reassessed QESE~\cite{wang2014psychometric} and the three ratings of perceived learning experience used in the pre-study survey, but targeted for \sysname{}, and was supplemented with follow-up questions probing the reasons for participants' ratings.
It also asked participants about their willingness to use \sysname{} on a 5-point Likert scale and included open-ended questions regarding their overall user experiences and suggestions for design improvements.
\idx{6}
\revised{For the post-study English proficiency test, we used the same question set as in the pre-study test, with both the question order and the answer-option order randomized to mitigate test–retest bias. Participants were not shown the correct answers after the pre-study test, ensuring that they could not learn directly from the test itself. Using identical questions is a common method for controlling variation in question-set difficulty (\cf{},~\cite{leong2024putting, inie2021aiki, wang2019minddot, hautasaari2019vocabura, kovacs2015feedlearn}). The three-week deployment period between the pre- and post-tests also provided a sufficiently long interval to minimize memory and practice effects~\cite{ebbinghaus2013image}.
}

\subsection{Expert Evaluation of Questions}
To assess the performance of the question generation pipeline (~\autoref{sec:quality}), we conducted an expert evaluation using a subset of questions generated during the deployment study.
\idx{14}\revised{We randomly sampled 30 questions: 24 in the question pool (\ie{}, those which passed the quality checking) and 6 that were eventually discarded due to unmet answerability or proficiency criteria.
17 of 24 questions in the pool had been presented to users during the study.}
\revised{To enable a direct comparison with the question generation pipeline, we applied the same criteria used by \sysname{}'s Question Evaluator (\cf{}, \autoref{fig:pipelines:generation}-\circledigit{h})---answerability and proficiency. Informed by prior work on educational question quality evaluation~\cite{elkins2024teachers,doughty2024comparative}, we developed a rubric comprising three items: one directly aligned with answerability and two aligned with proficiency. These items collectively operationalize our two evaluation criteria (see \autoref{appendix:expert_rubrics} for the full rubric and mappings).}

We recruited three English educators (E1--E3; all female) through social media advertisements.
Their professional backgrounds included university and high school teaching, as well as the development of standardized English test.
They were aged 39, 53, and 39, with 10, 20, and 15 years of experience in English education, respectively. \idx{14}\revised{The experts participated in remote evaluation sessions via Zoom. In the session, experts first evaluated the 30 questions through an online survey. They then went through a follow-up interview, where we asked about the potential benefits and concerns of the approach we take in \sysname{}, specifically generating EFL learning materials from their work context.}
The evaluation and interview took 90 minutes to complete. We compensated them with 100,000 KRW (approx. 72 USD).
\subsection{Data Analysis}
To examine participants’ engagement and usage patterns with \sysname{}, we conducted descriptive analyses of interactions with both \queryapp{} and \quizapp{}. For \queryapp{}, we analyzed message–response pairs, usage days and patterns, and the distribution of query prompts. For \quizapp{}, we examined the number of solved questions, usage days, and solving patterns, and quiz progress across repeated attempts.

To evaluate the generated questions, we examined pipeline performance and human evaluations. 
Pipeline performance was evaluated on 30 generated questions by comparing its binary judgments with expert labels. Ratings from three experts were aggregated by majority vote for answerability and proficiency, coded as true or false (with ``unknown'' mapped to false), and used as ground truth.
We then calculated precision, recall, and F1-score, with precision measuring agreement on pipeline-accepted items, recall measuring agreement on expert-accepted items, and F1 as their harmonic mean.
\idx{14}\revised{We transcribed the audio-recordings of the follow-up interviews with experts and conducted a qualitative analysis. One researcher coded the data using initial themes informed by the interview guide. The full research team then refined these themes through multiple rounds of peer debriefing, which surfaced the following themes: (1) the quality of AI-generated questions, (2) differences between \sysname{} questions and standardized English proficiency tests, and (3) considerations in English question design.}

We analyzed pre- and post-study surveys and tests to assess the effects of \sysname{} on participants’ learning performance and experiences.
To examine changes in learning performance over the study period, we analyzed the pre- and post-study English proficiency test using a mixed-effects model and the English self-efficacy questionnaires (QESE) using paired-samples t-tests.
\idx{5}\revised{The QESE met the normality assumption (Shapiro–Wilk test, $W=0.95$, $p>0.05$), allowing for parametric comparisons~\cite{hanusz2016shapiro}.}
To investigate learning experiences relative to prior practices, we compared participants’ perceived effectiveness of \sysname{} with their past EFL learning using the Wilcoxon signed rank test across three criteria: relevance of materials to work, helpfulness for actual work tasks, and sustainability of engagement.

\idx{14}\revised{To add more nuances to the quantitative findings in describing participants' experience and perceptions, we grouped participants' answers to the open-ended questions from the post-study surveys according to the following aspects: (1) the quality and relevance of generated questions, (2) effects of \sysname{} on learning and self-efficacy, and (3) the perceived benefits and drawbacks of \sysname{}. We incorporate this information in different sections of findings.}

%% file: tables/table-participants.tex
\begin{table}[t]
\sffamily
\def\arraystretch{1.15}
\setlength{\tabcolsep}{0.25em}
\centering

\caption{Demographic information and self-reported CEFR levels of the participants in our deployment study.}
\Description{Table listing 28 participants with pseudonyms, showing their age, gender, CEFR level, and job titles. Participants span diverse professions, with ages ranging from mid-20s to late-40s and CEFR levels from A1 to C1.}
\label{tab:participants}

\begin{tabular}{c c c c l}
\toprule
\textbf{Alias} & \textbf{Age} & \textbf{Gender} & \textbf{CEFR} & \textbf{Job title} \\
\midrule
\textbf{P1}  & 28 & Female & C1 & Global Business Developer \\
\textbf{P2}  & 28 & Female & A2 & Software Engineer \\
\textbf{P3}  & 39 & Female & B2 & Hospital Operations Manager \\
\textbf{P4}  & 28 & Male   & A2 & Medical Resident \\
\textbf{P5}  & 32 & Female & B1 & Product Designer \\
\textbf{P6}  & 36 & Female & B2 & Software Engineer \\
\textbf{P7}  & 47 & Female & B1 & Kindergarten Counselor \\
\textbf{P8}  & 45 & Female & B2 & Administrative Coordinator \\
\textbf{P9}  & 37 & Female & B2 & Office Manager \\
\textbf{P10} & 29 & Male   & A2 & International Patient Coordinator \\
\textbf{P11} & 30 & Male   & B2 & Sports Event Manager \\
\textbf{P12} & 38 & Female & B2 & Export--Import Specialist \\
\textbf{P13} & 42 & Male   & C1 & Professor \\
\textbf{P14} & 35 & Female & B2 & Apparel Export Manager \\
\textbf{P15} & 29 & Male   & A2 & Clinical Psychology Trainee \\
\textbf{P16} & 41 & Male   & A1 & IT Security Manager \\
\textbf{P17} & 27 & Female & B1 & Graduate Student \\
\textbf{P18} & 28 & Male   & B1 & Machine Learning Engineer \\
\textbf{P19} & 25 & Male   & B2 & Graduate Student \\
\textbf{P20} & 29 & Female & C1 & Real Estate Professional \\
\textbf{P21} & 32 & Female & A1 & Biotech Researcher \\
\textbf{P22} & 48 & Female & C1 & Logistics Specialist \\
\textbf{P23} & 25 & Female & B1 & Graduate Student \\
\textbf{P24} & 30 & Female & B2 & Marketing and Project Manager \\
\textbf{P25} & 33 & Male   & A1 & Network Engineer \\
\textbf{P26} & 31 & Female & C1 & Nuclear Policy Researcher \\
\textbf{P27} & 32 & Female & C1 & Sports Event Manager \\
\textbf{P28} & 30 & Male   & C1 & Governance Administrator \\
\bottomrule
\end{tabular}
\end{table}

%% file: sections/06_results.tex
\section{Findings}
\cameraready{In this section, we present findings from our deployment study in three parts. In \autoref{sec:results:engagement}, we provide an overview of the usage patterns of \sysname{} and report on participants' self-report sustainability (RQ1). In \autoref{sec:results:evaluation}, we investigate the quality of generated quizzes and how the use of \sysname{} supported participants' English proficiency and self-efficacy (RQ2). Lastly, in \autoref{sec:results:reactions}, we report on participants' perceived utility of \sysname{} and summarize their feedback on the strength and drawback of our approach (RQ3).} 

\subsection{\revised{Usage Patterns and Sustained Engagement}}\label{sec:results:engagement}
The interaction logs and usage data indicated that participants actively engaged with \sysname{}, frequently using both \queryapp{} and \quizapp{}. Here we report the descriptive statistics regarding participants' usage patterns and engagement with the two apps.

\subsubsection{Active Querying with \queryapp{}}
Across three weeks, participants opened a total of 652 conversation threads, and submitted 3,325 messages ($M=118.8$ per participant) through \queryapp{}.
On average, participants used \queryapp{} for 13.2 days ($SD=2.5$, $min=10$ [P15], $max=19$ [P5]), exceeding the required 10 days of use.
This indicates that participants engaged on most weekdays.
\revised{\autoref{fig:hourly_lingoquery}} presents participants' hourly engagement patterns, showing peak usage during work hours, particularly around 17 o'clock. 

Participants actively used pre-defined query prompts---\ie{}, \textit{Look-up}, \textit{Translate}, and \textit{Proofread}--- or wrote their own when submitting a message. 
Out of the 3,325 query-response pairs, \textit{Translate} responses were the most common (1,271 reponses; 38.2\%), followed by \textit{Look up} (399 responses; 12.0\%) and \textit{Proofread} (287 responses; 8.6\%).
The rest of the responses (1,369 responses; 41.2\%) were plain text messages, such as responses to queries asked in plain text or follow-ups.

\begin{figure}[t]
    \centering
    \includegraphics[width=\linewidth]{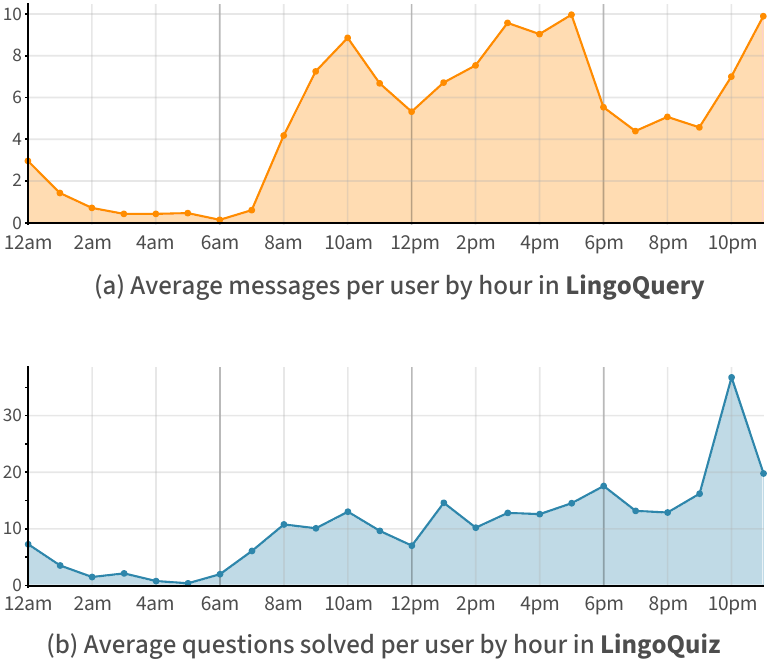}
    \caption{Hourly engagement patterns of \sysname{} during the three-week deployment. The orange line plot shows the average number of user messages in \queryapp{} per hour across a 24-hour day, with notable peaks around 10 a.m., 5 p.m., and 11 p.m. The blue line plot shows the average number of solved questions per user per hour in \quizapp{} across a 24-hour day, with a clear peak around 10 p.m.}
    \Description{An orange line plot showing hourly engagement with LingoQuery during the three-week deployment. The x-axis covers 24 hours of a day and the y-axis shows the average number of user messages per hour. Clear peaks appear around 10 a.m. (morning work hours), 5 p.m. (late afternoon), and 11 p.m. (late evening), while activity is lowest in early morning. A blue line plot showing hourly engagement with LingoQuiz during the three-week deployment. The x-axis covers the 24 hours of a day and the y-axis shows the average number of solved quiz questions per hour. Engagement is relatively steady during the day and reaches a distinct peak at 10 p.m., with lower activity during early morning hours.
}

    \label{fig:hourly_lingoq}
    \labelphantom{fig:hourly_lingoquery}
    \labelphantom{fig:hourly_lingoquiz}
\end{figure}

Regarding features, eighteen Participants regularly used the \textit{marking} feature \starmark, which ensures that the particular message pairs would appear in future quizzes. They marked 13.4 AI responses per person on average ($SD=16.1$). Of the 241 marked messages, 91 messages (37.8\%) were responses for \textit{Look up} queries, indicating participants' desire to revisit vocabulary or expressions.
Participants opened \queryapp{} by directly capturing the selected text and surrounding context using keyboard shortcuts \shortcut{}, for 6.9\% of all queries. However, only three participants (P11, P19, P26) dominated the usage of this feature and accounted for 65.8\% of all shortcut-triggered messages. 


\begin{figure*}[b]
     \centering
     \includegraphics[width=\textwidth]{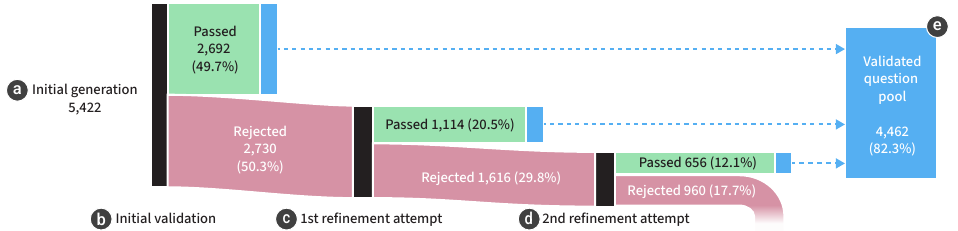}
     \caption{\revised{Overview of the question generation and validation to final question pool in the deployment study, starting from initially generated 5,422 questions \circledigit{a} to the 4,462 validated questions \circledigit{e} after three-stage validations with two refinements. The black vertical bars (\circledigit{b}, \circledigit{c}, \circledigit{d}) denote question evaluation steps.}}
     \Description{This figure shows the pipeline for generating and validating questions in the study. It begins with 5,422 generated questions (step a), followed by three evaluation stages (steps b, c, d) with two refinement cycles. After validation, 4,462 high-quality questions remained (step e).}
     \label{fig:funnel_question_validation}

 \end{figure*}

\subsubsection{\revised{Question Refinement and Validation}} \revised{Of the 3,325 query-response pairs, our question generation pipeline classified 2,711 (81.5\%) pairs as English language queries suitable for quiz generation, whereas 614 pairs (18.5\%) were excluded because they were not English-related queries.
Starting from the 2,711 English query pairs, the pipeline first generated 5,422 questions---twice the number of input pairs.
\autoref{fig:funnel_question_validation} summarizes the validation and refinement steps starting from these 5,422 initial questions (\circledigit{a} in \autoref{fig:funnel_question_validation}). After the initial validation (\circledigit{b} in \autoref{fig:funnel_question_validation}), 2,692 (49.7\%) questions passed the evaluation. 1,114 (20.5\%) questions passed on the second attempt after one refinement (\circledigit{c} in \autoref{fig:funnel_question_validation}), and 656 (12.1\%) passed after two refinement iterations (\circledigit{d} in \autoref{fig:funnel_question_validation}).
After these refinements, 960 (17.7\%) questions did not satisfy the evaluation criteria and were filtered out.
As a results, 4,462 validated questions were added to the question pool (\circledigit{e} in \autoref{fig:funnel_question_validation}) over the three weeks. Of these, 3,290 were eventually exposed to participants on \quizapp{} (117.5 per participant).}


\subsubsection{Consistent Language Practice with \quizapp{}} Including the reappeared cases, participants solved a total of 7,155 questions (255.5 per participant). These questions were curated in 604 quizzes and participants completed most ones, leaving only 10 quizzes incomplete (1.7\%) throughout the study period. Over the three weeks, Participants completed at least one quiz in \quizapp{} for 13.4 days on average ($SD=2.8$, $min=10$ [P27], $max=19$ [P21]), indicating similar compliance with \queryapp{}. 
Participants completed an average of 1.04 quizzes per day ($min=0.32$ [P9], $max=2.11$ [P8]), spending about 9.3 minutes ($SD=3.0$) per quiz.
This result is more than twice the number of quizzes required to qualify for study completion; the minimum requirement was 10 days out of 21, at least one quiz per day, or roughly 0.5 quizzes per day on average.
\revised{\autoref{fig:hourly_lingoquiz}} presents participants’ hourly engagement patterns, showing a more than twofold increase in usage around 10 p.m.
This pattern aligns with trends observed in other popular mobile applications.

Of 3,290 unique questions, 927 questions (28.2\%) appeared more than once, with the most frequently reappeared item occurring 15 times. On average, each question appeared across 2.86 quizzes ($SD=1.48$).
As the questions were repeatedly presented, participants became more likely to answer them correctly. The average accuracy for questions presented for the first time was 82.6\%, which gradually increased to 89.5\% upon the second exposure in another quiz, and further to 92.4\% upon the third exposure.

\subsubsection{\revised{Sustained Engagement}}
The Wilcoxon signed-rank test showed that the participants rated the sustainability of learning with \sysname{} ($M=3.96, SD=0.88$) significantly higher than their prior EFL learning experiences ($M=2.96, SD =0.79$), $z=-3.47$, $p<0.001$ (see \autoref{fig:survey_subjective:sustainability}).

\begin{figure*}[b]
    \centering
    \includegraphics[width=1\textwidth]{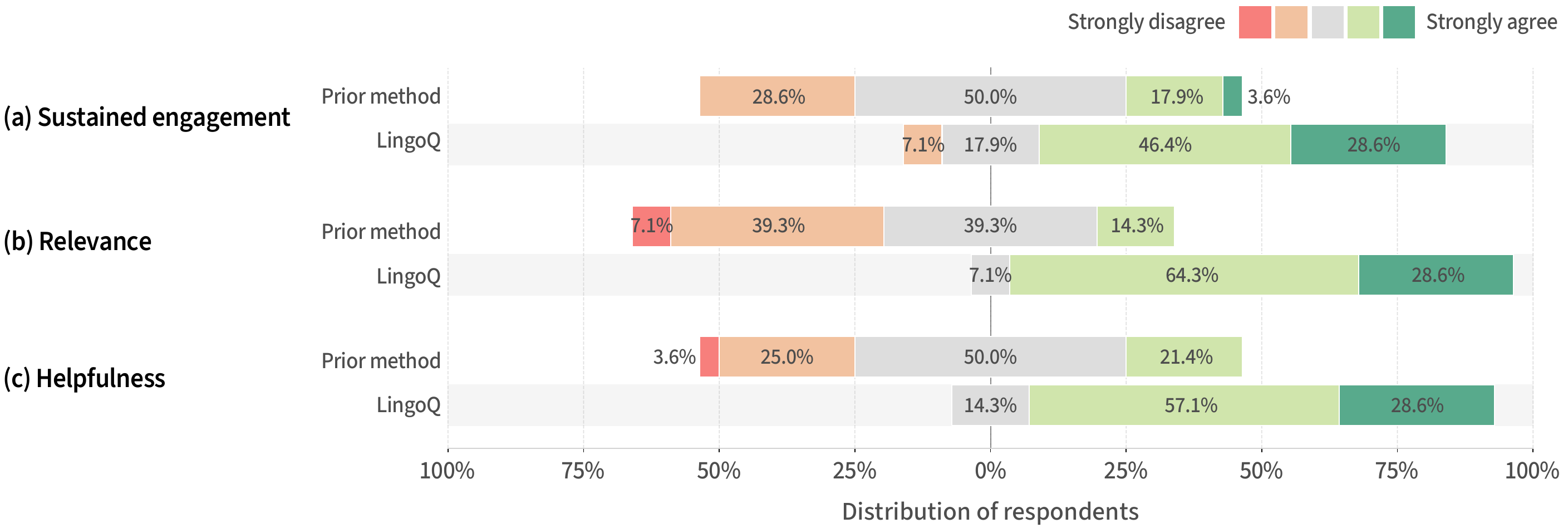}
    \caption{Stacked bar charts of five-point Likert ratings from participants ($N=28$) on (a) sustained engagement, (b) content relevance, and (c) helpfulness for work tasks, and  in learning. Upper bars indicate pre-study evaluations of existing EFL methods, while lower bars indicate post-study evaluations of \sysname{}.}
    \Description{Stacked bar charts of 28 participants’ Likert ratings on sustained engagement, content relevance, and helpfulness for work tasks. Upper bars show pre-study evaluations of existing EFL methods, and lower bars show post-study evaluations of LingoQ. Response categories are color-coded from left to right: red for strongly disagree, orange for disagree, gray for neutral, light blue for agree, and dark blue for strongly agree. A vertical center line marks the midpoint of the scale.}
    \label{fig:survey_subjective}
    \labelphantom{fig:survey_subjective:sustainability}
    \labelphantom{fig:survey_subjective:relevance}
    \labelphantom{fig:survey_subjective:helpfulness}

\end{figure*}

\subsection{\revised{Learning Experience Evaluation}}\label{sec:results:evaluation}
Over the three-week study, the use of \sysname{} significantly enhanced participants’ self-efficacy in using English at work. Proficiency gains varied by self-reported CEFR levels, with greater benefits for lower-level learners. These improvements were positively associated with \queryapp{} usage.

\subsubsection{\revised{Expert Evaluation}}
We compared the expert's assessments of the quality of genarated questions with the assessment from our automated quality assessment pipeline.
Our comparison yielded precision/recall/F1-scores of 0.91/0.81/0.86 for Answerability and 0.85/0.92/0.88 for Proficiency, suggesting that our automated filtering provided highly aligned decisions compared with the experts' judgment, with minor discrepancies. We identified two main reasons for the discrepancies.
First, experts often marked domain-specific questions as ``unknown'' (5 items for Answerability, 10 items for Proficiency), making it difficult to assess their quality as even the experts were not familiar with domain-specific terms.
Second, they applied a higher bar for proficiency, according to the follow-up interview, as they were accustomed to carefully adjusting difficulty to learners, to maintain discrimination of the test design.

In the follow-up interviews, all experts emphasized that the key difference between \sysname{} questions and standardized English tests is in the contexts used in the question stems, as E1 noted: ``\textit{TOEIC usually covers general business contexts, but some of the questions from \sysname{} required knowledge confined to highly specific domains.}''
Two experts (E2, E3) valued the domain-specific stems, noting that the learners' specialty in the domain could foster learning engagement and motivation, which are crucial factors for effective self-directed learning.
Meanwhile, E3 remarked that the overall quality of \sysname{} questions was comparable to text-completion items in TOEIC or TOEFL, noting that some items (\eg, \autoref{fig:question_examples:p11}) resembled high-quality questions that could plausibly appear on actual standardized English tests.

\subsubsection{English Proficiency}
\begin{figure}[t]
    \centering
    \includegraphics[width=\linewidth]{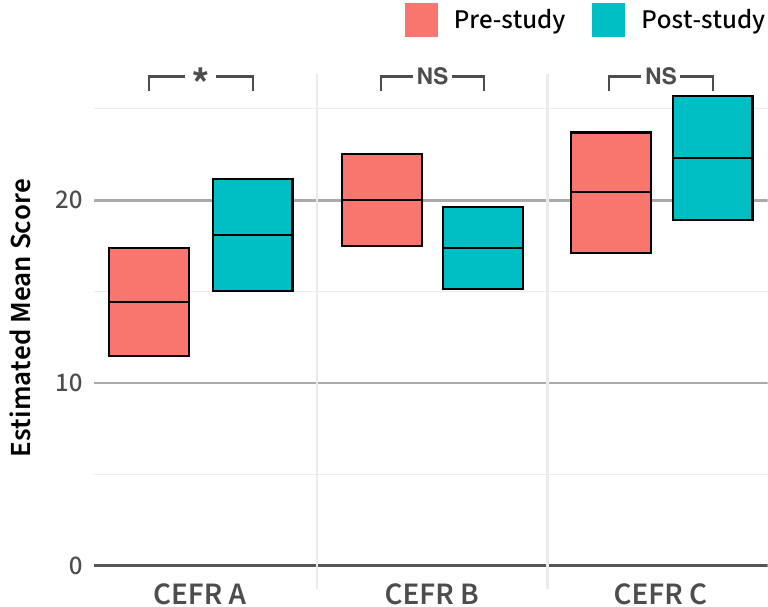}
    \caption{Estimated mean and 95\% confidence intervals of pre- and post-test English proficiency test scores by CEFR group. The plot shows estimated group means (with 95\% CIs) for \textit{A} (basic, $N=7$), \textit{B} (independent, $N=14$), and \textit{C} (proficient, $N=7$) on a 0–28 scale.}
    \Description{A plot of mixed effect model estimates for English proficiency scores ranging from 0 to 28. The x-axis shows pre- and post-test, and the y-axis shows test scores. Separate lines represent CEFR groups A (basic, N=7), B (independent, N=14), and C (proficient, N=7). The plot illustrates group mean scores with confidence intervals, allowing comparison of pre- and post-test performance across proficiency levels.}
    \label{fig:proficiency_test}
\end{figure}

A mixed-effects model analysis revealed a significant main effect of time on English proficiency scores, with an average increase of 1 point across all participants ($p=0.01$) (see \autoref{fig:proficiency_test}).
Post-hoc pairwise comparisons revealed that only \textsf{basic} (CEFR A) learners showed a significant improvement, gaining an average of 4 points (Total 30 points) from pre-to post-test ($p=0.01$), whereas \textsf{independent} (CEFR B) and \textsf{proficient} (CEFR C) participants showed no remarkable change.
However, for the \textsf{independent} group, the interaction between time and the number of \queryapp{} messages was significant ($p=.01$). 
The result indicates that more frequent use of the \queryapp{} was associated with greater learning gains among these EFL workers.
We further explain the learning effects of querying activity based on participants' general reactions in \autoref{sec:results:reactions}.

\subsubsection{English Self-Efficacy}
\begin{figure}[t]
    \includegraphics[width=\linewidth]{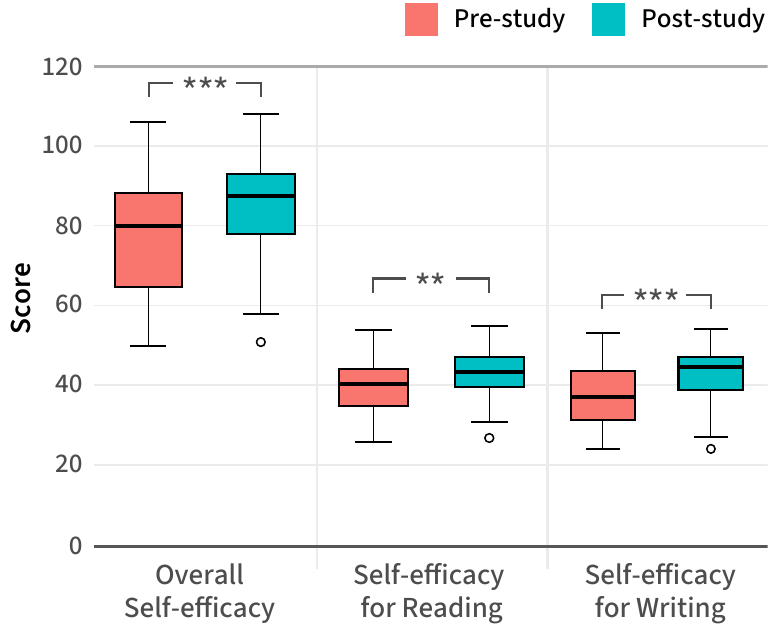}
    \caption{Box plots of English self-efficacy (QESE) scores on a 7-point scale. The left plot shows overall self-efficacy (16 items), while the middle and right plots show the subscales of reading (7 items) and writing (7 items). Significant pre–post differences are observed in both the overall scale and the subscales. Significance is marked as $p$ < 0.05 (*), $p$ < 0.01 (**), or $p$ < 0.001 (***).}
    \Description{Box plots of English self-efficacy (QESE) scores on a 7-point scale. Three plots are shown: overall self-efficacy with 16 items on the left, reading self-efficacy with 7 items in the middle, and writing self-efficacy with 7 items on the right. Pre-test scores are shown in blue and post-test scores in orange, with significant increases observed across all scales.}
    \label{fig:survey_qese}
\end{figure}

The paired $t$-test revealed  significant improvement in QESE score from pre- ($M=77.43$, $SD=14.31$) to post-study ($M=84.75$, $SD=13.21$) measurements ($t(27)=-4.30$, $p<0.001$, with 9.5\% gain (see \autoref{fig:survey_qese}).
In addition, both the reading and writing subscales of QESE also demonstrated significant gains ($t(27)=-3.67$, $p<0.01$ for reading, and $t(27)=-4.29$, $p<0.001$ for writing), respectively.
In the post-study survey, P15 highlighted the enhanced self-efficacy as the most notable benefit of \sysname{}, noting ``\textit{What improved the most was my confidence. It was really satisfying to go over the mistakes I often made, and over time, I found myself reading tough sentence structures much more easily.}''
Additionally, an expert from the expert evaluation reinforced this point: ``\textit{Confidence is a key factor in conversational ability, as it often translates into greater written and spoken outputs by reducing fear and hesitation. Thus, fostering confidence is essential for advancing from intermediate to higher proficiency levels.}'' (E1).



\subsection{\revised{Perceived Values of \sysname{}}}\label{sec:results:reactions}
In the post-study survey, we gathered participants' feedback on \sysname{} across multiple aspects. 
To gauge the utility of \sysname{} beyond the study context, we asked participants how much they would be willing to use \sysname{} in their real life (\sysname{} on their phones and computers would continue to work). 24 of 28 participants (86\%) indicated they would adopt the system, with 54\% selecting ``agree” and 32\% selecting ``strongly agree.”
We summarize their feedback on the strengths and drawbacks we identified.

\subsubsection{\revised{Content Relevance}}
The Wilcoxon signed-rank test revealed that participants rated the relevance of \sysname{} ($M=4.21, SD=0.57$) significantly higher than prior EFL practices ($M=2.61, SD=0.83$), $z=-4.39$, $p<0.001$ (see ~\autoref{fig:survey_subjective:relevance}).
Their open-ended response revealed that most participants (25 out of 28; 89.2\%) valued the quizzes for reflecting the practical English they actually encountered at work.
In particular, P15 emphasized the value of context-relevant words: ``\textit{I liked repeatedly practicing verbs specific to the medical field rather than casual spoken language.}'' (see \autoref{fig:question_examples:p15}).
Moreover, P11 and P20 found that domain-relevant distractors in quiz alternatives helped them contrast similar terms and deepen their understanding of subtle distinctions (see \autoref{fig:question_examples:p20}).

\subsubsection{\revised{Helpfulness for Work Tasks}}
The Wilcoxon signed-rank test revealed that participants rated \sysname{} ($M=4.14, SD=0.65$) as more helpful for daily work tasks than their prior EFL practices ($M=2.89, SD=0.79$), $z=-4.05$, $p<0.001$ (see \autoref{fig:survey_subjective:helpfulness}).
In the post-study survey, participants reported that practicing work-related content with \quizapp{} not only improved retention but also made English-related work tasks smoother and more efficient.
They remarked that \quizapp{} reinforced their learning, allowing them to ``\textit{encounter the content again through quizzes} (P17).
In particular, P18 highlighted that reviewing previously read content through quiz questions enhanced their reading fluency, noting, ``\textit{Since I had the opportunity to revisit documents and papers I had read before, I found that when rereading, I was able to process them more quickly in English.}'' (P18).

\subsubsection{Expanding Types of Questions and Language Learning Disciplines}
Most participants (24 out of 28; 85.7\%) found the quiz design effective after work.
\idx{1,15}\revised{P15 emphasized that \sysname{} enabled effortless learning through highly context-relevant materials, noting that ``\textit{Before this, I used Gemini with my roommate to study vocabulary, but we always had to manually instruct what to practice [...]
LingoQuiz removed that burden.}''}
In addition, P3 noted, ``\textit{The quizzes weren’t burdensome and fit easily into my routine, like during commutes or before bed.}'', and, P20 remarked, ``\textit{I could learn just by doing the quizzes without the extra step of studying beforehand.}''.

\revised{Although the lightweight quiz format helped sustain review routines, nine} participants suggested diversifying the quiz formats beyond the current fill-in-the-blank design. They proposed that exercises could be more closely aligned with the types of queries submitted. For instance, for translation queries, quizzes could present multiple sentence options and ask learners to select the correct translation. 
In addition, five participants suggested expanding the material modalities to include speaking and listening practice, aiming to better support verbal communication tasks such as video meetings and presentations.

\subsubsection{\idx{11,15}\revised{Utility of Tailored Interaction Components}}
\revised{Around half of participants (13 out of 28) perceived \queryapp{}'s bespoke interaction features tailored to English queries---such as quick-access buttons for predefined query intents and keyboard shortcuts---to be particularly convenient and useful. For example, P12 remarked, ``\textit{The three template prompt buttons were useful because I didn’t have to keep typing the same prompts.}''
However, three participants also reported mixed experiences with this language-specialized design.
While participants valued the tailored linguistic support, they found the system limiting when they needed assistance with features typically supported by general-purpose LLMs (e.g., file upload).
P5 noted, ``\textit{I ended up keeping another AI tool open alongside \sysname{} while working.}''}

\subsubsection{Perception Change of Queries as Learning Opportunities}
While we provided \queryapp{} as a dedicated channel for language querying, three participants mentioned that using such a scoped interface raised awareness of knowledge gaps and encouraged reflection on their English use when they ask questions. They contrasted this experience with their prior experience with AI assistants. P25 remarked, ``\textit{Using \queryapp{} instead of ChatGPT helped me develop the habit of looking more carefully at words in sentences I would have otherwise translated without much thought.
}'' 
Knowing that their queries would generate learning materials for later completion made participants see each query as part of their language learning. As a result, using \queryapp{} reminded them of EFL learning even when their questions were unrelated. 
These reflections suggest the potential to reorient reliance on LLMs toward active learning.

\subsubsection{Backfire of Authentic Materials}
While participants valued the activity of solving work-related quizzes, two also noted that practicing quizzes containing work-related materials after work sometimes depressed them.
P8 noted, ``\textit{Sometimes I wanted to detach from work, but reviewing the same materials after hours felt like an extension of my job.}''
\idx{15}\revised{Four participants mentioned that although \sysname{}’s high level of personalization was helpful, it occasionally felt overly tied to their own queries.
They pointed out that the system generated quizzes based on explicit queries, which limited broader learning opportunities. 
P1 explained, ``\textit{To learn related words, I had to explicitly ask \queryapp{}. For example, when I came across `customer churn' in context, I needed to ask follow-up questions like `How is it different from customer retention?' for those to appear in the quiz.}''
P19 suggested augmenting the content with paraphrased alternatives or domain-specific vocabulary that the system could infer as relevant, even without explicit learner requests.
}

%% file: sections/07_discussion.tex
\section{Discussion}

Our results highlight how \sysname{} bridges two familiar practices---using AI tools at work for English-related tasks and studying English on smartphones—by turning routine queries into learning activities that are directly connected to workers’ tasks and that strengthen their self-efficacy. These findings inform design implications for work-integrated language learning systems that respect workers' boundaries and ethical considerations.

\subsection{Leveraging Reliance on LLMs for Learning Opportunities}

Reliance on generative AI has been a threat to learning as it reduces an opportunity for a critical engagement with a subject matter~\cite{chan_ai_2023}. Especially for workers, the convenience that LLMs provide fosters passive consumption of generated information rather than critically examining what they produce or comprehend~\cite{10.1145/3706598.3713778}. 
Therefore, ironically, EFL workers' reliance on convenient LLM-based tools can lead to the deterioration of their English skills. 

\idx{1} 
\revised{In this work, we leveraged the conversational data that people generate while interacting with an LLM-based tool. Although such data is often used to enhance conversational quality within a session and personalize future interactions~\cite{liu-etal-2024-lost}, or can be explicitly retained when users opt in to maintain personal memory~\cite{bae2022keep}, reusing and managing this stored information remains challenging for end users~\cite{10.1145/3654777.3676388, 10.1145/3613904.3642754, 10.1145/3544548.3580817}. Theoretically, a worker could generate learning materials directly from the LLM's memory (\eg{}, talking to the assistant, ``\textit{Based on the conversation history, generate fill-in-the-blank English questions that will help me improve the work-related English skills I need.}''). However, our results indicate that simple prompting does not guarantee the validity of the generated questions; more than 50\% of the generated questions initially were either insufficiently proficient or unanswerable (see \circledigit{b} in \autoref{fig:funnel_question_validation}). Moreover, the generation process to support effective learning---such as producing varying questions, marking vocabulary that they want to study, or revisiting items previously answered incorrectly---would require substantial manual effort or additional technical development. 
As a result, participants perceived their LLM queries at work not merely as assistance but as opportunities for language learning. Moreover, although we primarily focused on piggybacking~\cite{grevet2015piggyback, epstein2022piggyback} on workers' existing interaction behaviors with AI assistants when designing \queryapp{}, participants} indicated that the tailored UI features---such as side-by-side translation view, toggling between refined and original responses---helped them become aware of what they did not know and facilitated conscious learning.
The noticing hypothesis~\cite{robinson1995attention} suggests that conscious awareness of linguistic gaps is essential for acquisition, beyond mere exposure. 
Unlike passive reliance on AI-generated answers, this awareness might have reframed their work-related queries as active learning events.



\revised{\idx{} Due to the short span of the study, it remains unwarranted if the system would elicit sustained engagement for a longer term.
The predictable one-to-one mapping between queries and questions may hinder sustained engagement for the limited varied practice~\cite{schmidt1992new} or a desirable difficulty~\cite{bjork2011making}, which are essential for retention and transfer of vocabulary knowledge~\cite{bjork2015desirable}.} One direction to diversify quiz content is to evaluate a user's proficiency level informed by user modeling based on the collection of query-response pairs~\cite{ding2025unknown,chen2024worddecipher,higasa2024keep,arakawa2022vocabencounter,lungu2018we}.
Varying the stem for creating a new scenario that uses the same words and expressions can mitigate the reviewing, not anticipating, nature of \sysname{}~\cite{peng2023storyfier,yamaoka2022experience}.


\subsection{\revised{Meaningful Increases in Self-Efficacy Despite Partial Proficiency Gains}}
\idx{9} \revised{Using \sysname{} led to a significant increase in participants' self-efficacy in English, while measurable learning gains were only observed among the basic group. The result offers a promising indication of \sysname{}'s long-term potential. A substantial body of work in second-language acquisition identifies self-efficacy as one of the strongest predictors of future learning gain; multiple literature reviews show a positive association of self-efficacy with L2 learning outcomes and proficiency levels~\cite{Sun2021Relationship, Goetze2022Is}. This boost suggests that \sysname{}’s impact may extend beyond the study window, offering promising long-term learning potential.} 

\idx{9}\revised{Several factors may explain why participants' higher proficiency groups did not exhibit learning gains. Previous work shows that early vocabulary development yields rapid, detectable improvement, whereas intermediate and advanced learners experience diminishing returns because additional vocabulary is less frequent, harder to acquire, and contributes minimally to standard proficiency measures~\cite{schmitt2014reassessment}. Given that our fill-in-the-blank questions primarily targeted vocabulary and expression, the level of the generated items may have been too easy for intermediate and advanced learners.}

\idx{9} \revised{
Users’ goals, which vary by proficiency level, may also have influenced learning outcomes. Basic-level participants likely relied on \queryapp{} because they genuinely did not know word meanings or were unable to translate. In contrast, intermediate and advanced users may have turned to \queryapp{} not out of incapacity but to speed up their work. This pattern aligns with prior observations in software developers, who use automation to offload trivial or repetitive tasks~\cite{mohamed2025impactllmassistantssoftwaredeveloper}. Advanced users may similarly leverage LLMs to optimize workflows (e.g., drafting an email), even when the query provides little new learning content. Consequently, \quizapp{} may have been less effective for higher-level participants, as the generated questions often covered material they had already mastered.}

Future systems should consider how generative approaches can better support learners across a broader proficiency spectrum. \idx{12}\revised{Distinguishing the intent behind each query---those driven by knowledge gaps (\eg{}, looking up a word in a dictionary) versus those made for efficiency (\eg{}, translating text into English)---may enable targeted scaffolding.} 
Such data can be further used to create a learner profile \idx{12}\revised{and to provide additional context for adaptive question generation, enabling more robust proficiency estimation. 
This, in turn, would allow systems to generate questions with desirable difficulty levels and targeted style (\eg{}, summarization, paraphrasing, error correction) that promote progression even for advanced learners. 
}

\subsection{\revised{Respecting Privacy in Work-Related Learning Systems}} \idx{16} \begin{revisedenv}\sysname{} allowed users to include a screenshot to augment question generation. Although we allowed opting out (\ie, discard the screenshot before sending the message), taking a screenshot may be against the user's company's security policy, thereby creating a risk of unintentional violation. Moreover, screenshots may also include confidential personal/workplace content, posing significant privacy risks if such data is leaked. Therefore, it is critical that \sysname{} provide users with clear awareness of and control over what data is captured and how it is used, enabling contextualized learning without placing them at risk of personal or professional harm.
\idx{16} \revised{Recent developments in edge computing and on-device lightweight vision–language models offer promising pathways toward privacy-preserving alternatives~\cite{11032085, xu2024device}. Future implementations could process screenshots locally or on edge devices rather than sending them externally, enabling contextual personalization while reducing the exposure of sensitive workplace content.}
\end{revisedenv}


\subsection{\revised{Balancing Learning and Detachment in Workplace Contexts}}
\idx{9, 16} \revised{Using \sysname{} may also influence how workers negotiate the boundaries between work and personal life. Because the system generates learning materials directly from workplace contexts, it can blur the boundary between work and life. In the post-study survey, some participants expressed concern that \sysname{} could make it harder to fully detach from work, creating subtle pressure to engage with work-related content, which could potentially affect their mental well-being~\cite{binnewies2010recovery, sonnentag2015recovery, cropley2015relationship}. 
}

\idx{9, 16} \revised{These concerns highlight the need for responsible design in AI-powered personalized learning, especially when linking personal data to personal development. Future systems should preserve users’ agency by giving them control over when learning materials appear, keeping engagement optional, and avoiding content that adds stress. Such a design supports voluntary, time-bounded participation and reduces the blurring of work and life. More intelligent approaches could also retain the English content users need to study while altering its surrounding context, minimizing reminders of workplace tasks and reducing the sense of continuous work exposure.}

\subsection{Limitations and Future Work}

In this section, we discuss the limitations of this study. 
We focused on reading and writing skills, whereas participants expressed a need for support in listening and speaking. This suggests that future systems should incorporate speaking and listening to better support verbal communication. Similarly, the problem format was limited to fill-in-the-blank questions, which might have limited the learning effects of the system.

Our evaluation employed a single-group study design, focusing on the feasibility and understanding of engagement with the system deployed in the field. While this design allowed us to observe real-world usage and effects, future work could incorporate comparative study designs to attribute the effects of work-related English learning to \sysname{}'s connected pipeline.
\cameraready{Additionally, we imposed a minimum usage requirement to mitigate noise from non-usage attrition. Although this threshold helped ensure data quality for behavioral analysis, it might also have influenced the observed engagement levels; therefore, findings related to sustained engagement (RQ1) should be carefully interpreted.}


Third, our study was limited to a Korean–English context. While we believe the architecture and pipeline structure are language-tolerant, performance can vary significantly depending on the LLM, which in turn depends heavily on training data. Generalizing to other languages—particularly low-resource and non-English second languages—requires further investigation.

%% file: sections/08_conclusion.tex
\section{Conclusion}

We presented \sysname{}, an LLM-based system that supports learning work-related English skills by generating quizzes directly from workers’ language queries to LLM tools. By connecting English query history to low-burden practice, \sysname{} enables work-related English exercises anytime and anywhere. \cameraready{To answer our three research questions,} we conducted a three-week deployment with 28 EFL information workers. Participants actively engaged with the system and reported \cameraready{more sustainable review practices compared to their previous EFL learning methods (RQ1). 
Our study revealed that queries can be effectively transformed into learning materials 
of sufficient quality---as confirmed by expert evaluation---which in turn led to increased self-efficacy for all participants and measurable gains for beginners, with further potential for advanced learners through more active system interaction (RQ2). 
Overall, participants valued \sysname{} as more contextually relevant and helpful for work than their prior English study methods (RQ3).}
These findings demonstrate how leveraging workers’ reliance on LLMs can create new opportunities for personalized learning, while still respecting work boundaries and ethical considerations. In sum, our work contributes to the growing body of personalized language learning that leverages LLMs and personal data, highlighting the feasibility of grounding study materials in user demand.

%% file: appendix/prompts.tex
\section{LLM Instruction Prompts Used for \queryapp{}}

\subsection{Instructions for Conversational Agent Intent Classifier}\label{appendix:prompts:query:intent}
\begin{lstlisting}[language={}, basicstyle=\ttfamily\small]
[Role]
You are an Intention Classifier. Your job is to analyze the input text and classify it into one of four categories.

[Classification Categories]

**translation**: "Translate the following text naturally between English and Korean. Please also explain how the nuance and context of the sentences are reflected in the translation."

**proofread**: "Proofread the following text into more accurate and natural English. Please also provide an explanation of the changes and the reasons behind them."

**lookup**: "Explain the meaning of the following word (or expression) in detail, in the style of a dictionary entry."

**text**: Any other input that doesn't match the above three categories.

[Classification Rules]
1. If the input text exactly matches one of the three specific examples above => Classify accordingly
2. If the input text is similar to any of the three examples => Classify accordingly
3. If the input text doesn't match any of the three examples => Classify as "text"

[Output Format]
**CRITICAL: You must respond with ONLY ONE WORD from the list below.**
**DO NOT use JSON format. DO NOT add explanations. DO NOT add quotes.**

Respond with ONLY one of these four values:
- translation
- proofread
- lookup
- text 
\end{lstlisting}

\subsection{Instructions for Conversational Agent Response Generator}\label{appendix:prompts:query:response}
\begin{lstlisting}[language={}, basicstyle=\ttfamily\small]
[Role]
You are a Workplace English Support Assistant, designed to help the user tackle English-related tasks and challenges in everyday work situations.

[Personality]
- Patient and encouraging
- Clear and articulate in explanations
- Friendly and approachable
- Professional yet conversational
- Culturally sensitive and inclusive

[Chat Style]
- The user will speak in {user_language}. So you must also speak in polite and supportive {user_language}.
- Do not greet the user and treat them as if you already know them well.

[CRITICAL: Context Memory & Style Consistency]
- ALWAYS remember the entire conversation history
- Remember user's work context, preferences, and instructions
- Remember user's ongoing projects and tasks
- Maintain consistent response style throughout the conversation
- If user prefers certain response styles, maintain that consistency

[Message Content Format]
The user's intention is provided as: [Intention: INTENTION_PLACEHOLDER]
The user's message contains:
- query_prompt: The prompt the user is using to make the query
- content: The content the user is querying about

[Intent-Based Response Generation]
IMPORTANT: The user's intention has already been classified. Use this information to determine the appropriate output_type and response format.

**Response Format Based on Intention:**

1. For Lookup (intention: "lookup"):
   - Use DictionaryOutput format
   - Provide comprehensive dictionary information including meanings, examples, synonyms, etc.
   - Focus on the word/phrase in the user's content

2. For Translate (intention: "translation"):
   - Use TranslationOutput format
   - Provide original text, translation, and explanation
   - Translate naturally, considering user's context and communication style
   - Avoid literal translation - focus on natural expression
   - **Pay attention to formality, tone, and context**: Match the user's professional level, industry terminology, and communication style
   - When the user content is a mix of {user_language} and English, translate the entire content into English

3. For Proofread (intention: "proofread"):
   - Use RefinementOutput format
   - Provide original content, refined content, and refinement rationale
   - Refine naturally
   - **Minimal refinement approach**: Preserve the user's original structure and meaning as much as possible.
   - **refinement_rationale**: Write in simple, natural Korean. Avoid numbered lists or structured formats.
   - When the user content is a mix of {user_language} and English, refine the content to be fully in English
   - Only refine to {user_language} if the user explicitly requests it

4. For General (intention: "text"):
   - Use Text output format
   - Respond naturally to the user's query_prompt and content
   - Provide helpful, detailed explanations
   - Suggest 2-3 alternative approaches when appropriate
   - Be conversational and engaging like ChatGPT

**Your Task:**
Based on the classified intention provided, generate the appropriate response using the correct output_type and format. Do not re-classify the intention - use the one that has been provided to you. 
\end{lstlisting}

%% file: appendix/quizevaluation.tex
\section{Development and Validation of the English Proficiency Test}\label{appendix:test_validation}

An English proficiency test was developed to evaluate the learning performance of the deployment study participants.
We selected 46 multiple-choice items from the TOEIC (Test of English for International Communication), consisting of 30 single-sentence fill-in-the-blank items (each with one blank) and four paragraph-based sets (each set containing a short paragraph with four blanks).

\subsection{Participants}
To validate the difficulty and time required to complete the test, we recruited computer-based information workers via social media advertisements, following the inclusion criteria described in 
\autoref{sec:participants}.
Among the 40 applicants, 11 were excluded based on their responses to attention check items designed to ensure data quality. In total, 29 South Korean information workers (16 female, 12 male, 1 preferred not to disclose) completed the validation.
Participants had an average age of 27.9 years ($SD=4.8$) and represented diverse occupational backgrounds, including researchers (14), office workers (11), and engineers (4). Based on CEFR self-assessment~\cite{council2001common}, 2 participants identified as \textit{A1} (beginner), 2 as \textit{A2} (elementary), 8 as \textit{B1} (intermediate), 5 as \textit{B2} (upper-intermediate), 5 as \textit{C1} (advanced), and 7 as \textit{C2} (proficient). Each participant received 20,000 KRW (approx. 14 USD) as compensation.

\subsection{Procedure}
Participants completed the validation via an online survey. They solved all 46 test items along with 2 attention check questions. Problem-solving time was recorded. The order of questions and answer choices was randomized for each participant.
The average response time was 23.7 seconds for single-sentence items and 117.2 seconds for paragraph-based sets.
Mean scores were $M=21.03$ ($SD=4.88$) for the single-sentence items (score range: 0–30) and $M=11.24$ ($SD=2.43$) for the paragraph sets (score range: 0–16).

\subsection{Validation}
Based on classical test theory~\cite{devellis2006classical}, item difficulty was calculated as the proportion of participants who answered each item correctly. Following the standard range of acceptable difficulty (0.4 to 0.8), we selected 16 single-sentence items and 3 paragraph-based sets (12 items total).
Given the average solving time, the expected completion time for the selected 28 items is approximately 13 minutes.
Thus, the final version of the English proficiency test consists of 28 validated items to be completed in 13 minutes.

%% file: appendix/expertrubric.tex
\section{Expert Evaluation of Generated Questions}\label{appendix:expert_rubrics}
To evaluate the performance of the question evaluator in the \sysname{} pipeline, we conducted an expert evaluation on sampled 30 questions generated during the deployment study.
\revised{The expert evaluation rubric (\autoref{tab:rubric}) consisted of three items, which were mapped to two evaluation criteria.
Questions that satisfied the \textit{Correct answer} rubric were coded as \textbf{Answerability = True}, and \textbf{Proficiency = True} was coded only when both the \textit{Unique choices} and \textit{No obviously wrong} rubrics were satisfied.
}

\input{tables/table-rubrics}

%% file: tables/table-rubrics.tex

\begin{table*}[h]
\centering
\caption{Rubric used for expert evaluation of questions generated by \sysname{}.}
\Description{Table with three columns and three main rows. The columns list the rubric category, the evaluation question, and the answer options. Each row represents one rubric item used by experts to evaluate the quality of generated questions.}
\label{tab:rubric}
\small 
\renewcommand{\arraystretch}{1.15} 
\begin{tabularx}{\textwidth}{p{0.15\textwidth} p{0.35\textwidth} X}
\toprule
\textbf{Rubric} & \textbf{Question} & \textbf{Options} \\
\midrule
\multirow{5}{*}{Correct answer} 
& \multirow{5}{=}{Is there a correct answer listed in the options? 
Is the option marked “correct” actually correct?} 
& Yes, there is a correct answer and it is marked 'correct' \\
& & There is a correct answer but it is not marked 'correct' \\
& & There are multiple correct answers \\
& & No, there is no correct answer \\
& & Don’t know \\
\midrule
\multirow{4}{*}{Unique choices} 
& \multirow{4}{=}{Are the options distinct from each other, ensuring they are unique choices?} 
& Yes, they are completely unique \\
& & Some choices are unique, some are too similar \\
& & No, they are all too similar \\
& & Don’t know \\
\midrule
\multirow{4}{*}{No obviously wrong} 
& \multirow{4}{=}{Is the MCQ free from obviously-wrong options?} 
& Yes, there are no obviously-wrong options \\
& & Yes, but the options give away the correct answer \\
& & No, there are obviously-wrong options \\
& & Don’t know \\
\bottomrule
\end{tabularx}
\end{table*}

%% file: bibliography.bib
@string{retrieveddate = "Sep 1, 2025"}

@string{retrievedyear = 2025}

@misc{electron,
    title={{Electron: Build Cross-Platform Desktop Apps with JavaScript, HTML, and CSS}},
    url={https://www.electronjs.org/},
    author={{OpenJS Foundation and Electron contributors}},
    year= retrievedyear,
    lastaccessed = retrieveddate
}

@misc{openai,
    title={{OpenAI API}},
    url={https://openai.com/api/},
    author={OpenAI},
    year= retrievedyear,
    lastaccessed = retrieveddate
}

@misc{TypeScript,
  title        = {{TypeScript}},
  author       = {Microsoft},
  year         = retrievedyear,
  url          = {https://www.typescriptlang.org},
  lastaccessed = retrieveddate
}

@misc{FastAPI,
  title        = {{FastAPI Framework, High Performance, Easy to Learn, Fast to Code, Ready for Production}},
  author       = {FastAPI},
  year         = retrievedyear,
  url          = {https://fastapi.tiangolo.com/},
  lastaccessed = retrieveddate
}

@misc{LangChain,
  title        = {{LangChain: Applications that Can Reason}},
  author       = {LangChain, Inc.},
  year         = retrievedyear,
  url          = {https://www.langchain.com/},
  lastaccessed = retrieveddate
}

@misc{PostgreSQL,
  title        = {{PostgreSQL: The World's Most Advanced Open Source Relational Database}},
  author       = {The PostgreSQL Global Development Group},
  year         = retrievedyear,
  url          = {https://www.postgresql.org/},
  lastaccessed = retrieveddate
}

@misc{ReactNative,
    title={{React Native - Learn Once, Write Everywhere}},
    url={https://reactnative.dev/},
    author={Meta},
    year= retrievedyear,
    lastaccessed = retrieveddate
}

@misc{duolingomax,
    title={{Duolingo Max: AI-Powered Language Learning with GPT-4}},
    url={https://blog.duolingo.com/duolingo-max/},
    author={{Duolingo}},
    year= retrievedyear,
    lastaccessed = retrieveddate
}

@misc{googlelittlelessons,
    title={{Google Search: Little Language Lessons}},
    url={https://labs.google/lll/en/},
    author={{Google}},
    year= retrievedyear,
    lastaccessed = retrieveddate
}

@misc{duolingo,
    title={{Duolingo: Language Lessons for Everyone}},
    url={https://www.duolingo.com/},
    author={{Duolingo}},
    year= retrievedyear,
    lastaccessed = retrieveddate
}

@misc{babbel,
    title={{Babbel: Learn Languages Online}},
    url={https://www.babbel.com/},
    author={{Babbel}},
    year= retrievedyear,
    lastaccessed = retrieveddate
}

@misc{memrise,
    title={{Memrise: Language Learning Made Fun}},
    url={https://www.memrise.com/},
    author={{Memrise}},
    year= retrievedyear,
    lastaccessed = retrieveddate
}

@misc{rosettastone,
    title={{Rosetta Stone: Language Learning Software}},
    url={https://www.rosettastone.com/},
    author={{Rosetta Stone Ltd.}},
    year= retrievedyear,
    lastaccessed = retrieveddate
}

@misc{quizlet,
    title={{Quizlet: Learning Tools and Flashcards}},
    url={https://quizlet.com/},
    author={{Quizlet Inc.}},
    year= retrievedyear,
    lastaccessed = retrieveddate
}

@misc{anki,
    title={{Anki: Powerful, Intelligent Flashcards}},
    url={https://apps.ankiweb.net/},
    author={{Anki Developers}},
    year= retrievedyear,
}

@misc{chatgpt,
    title={{ChatGPT}},
    url={https://chat.openai.com/},
    author={{OpenAI}},
    year= retrievedyear,
    lastaccessed = retrieveddate
}

@misc{gemini,
    title={{Gemini}},
    url={https://gemini.google.com/},
    author={{Google DeepMind}},
    year= retrievedyear,
    lastaccessed = retrieveddate
}

@misc{claude,
    title={{Claude}},
    url={https://claude.ai/},
    author={{Anthropic}},
    year= retrievedyear,
    lastaccessed = retrieveddate
}

@misc{grammarly,
    title={{Grammarly: AI Writing Assistance}},
    url={https://www.grammarly.com/},
    author={{Grammarly Inc.}},
    year= retrievedyear,
    lastaccessed = retrieveddate
}

@misc{deepl,
    title={{DeepL Translator}},
    url={https://www.deepl.com/translator},
    author={{DeepL SE}},
    year= retrievedyear,
    lastaccessed = retrieveddate
}

@misc{ringle,
    title={{Ringle: 1:1 Online Tutoring with Ivy League Tutors}},
    url={https://www.ringleplus.com/},
    author={{Ringle English Education Service}},
    year= retrievedyear,
    lastaccessed = retrieveddate
}

@misc{speakapp,
    title={{Speak: AI-Powered English Tutoring App}},
    url={https://www.speak.com/?lang=en},
    author={{Speak Global Inc.}},
    year= retrievedyear,
    lastaccessed = retrieveddate
}

@book{lave1991situated,
  title={Situated Learning: Legitimate Peripheral Participation},
  author={Lave, Jean and Wenger, Etienne},
  year={1991},
  publisher={Cambridge university press}
}

@inproceedings{viberg2012mobile,
  title={Mobile Assisted Language Learning: A Literature Review},
  author={Viberg, Olga and Gr{\"o}nlund, {\AA}ke},
  booktitle={11th world conference on mobile and contextual learning},
  year={2012}
}

@book{nunan2004task,
  title={Task-Based Language Teaching},
  author={Nunan, David},
  year={2004},
  publisher={Cambridge University Press}
}

@book{Hutchinson1987ESP, 
place={Cambridge}, 
series={Cambridge Language Teaching Library}, 
title={English for Specific Purposes}, 
publisher={Cambridge University Press}, 
author={Hutchinson, Tom and Waters, Alan}, 
year={1987}, 
collection={Cambridge Language Teaching Library}}

@book{Willis2021TaskBasedLearning,
  title={A Framework for Task-Based Learning},
  author={Willis, Jane},
  year={2016},
  publisher={Longman}
}

@book{Coyle2010CLIL,
  title     = {CLIL: Content and Language Integrated Learning},
  author    = {Coyle, Do and Hood, Philip and Marsh, David},
  year      = {2010},
  publisher = {Cambridge University Press},
  isbn      = {9780521130219}
}

@article{long1996role,
  title={The Role of the Linguistic Environment in Second Language Acquisition},
  author={Long, Michael},
  journal={Handbook of Second Language Acquisition},
  pages={413--468},
  year={1996},
  publisher={Academic Press}
}

@article{cooke2010assessing,
  title={Assessing Concurrent Think-Aloud Protocol as a Usability Test Method: A Technical Communication Approach},
  author={Cooke, Lynne},
  journal={IEEE Transactions on Professional Communication},
  volume={53},
  number={3},
  pages={202--215},
  year={2010},
  publisher={IEEE}
}

@article{wang2014psychometric,
  title={Psychometric Properties of a Self-Efficacy Scale for English Language Learners in China},
  author={Wang, Chuang and Kim, Do-Hong and Bai, Rui and Hu, Jiyue},
  journal={System},
  volume={44},
  pages={24--33},
  year={2014},
  publisher={Elsevier}
}

@article{braun2006using,
  title={Using Thematic Analysis in Psychology},
  author={Braun, Virginia and Clarke, Victoria},
  journal={Qualitative Research in Psychology},
  volume={3},
  number={2},
  pages={77--101},
  year={2006},
  publisher={Taylor \& Francis}
}

@inproceedings{doughty2024comparative,
  title={A Comparative Study of AI-Generated (GPT-4) and Human-Crafted MCQs in Programming Education},
  author={Doughty, Jacob and Wan, Zipiao and Bompelli, Anishka and Qayum, Jubahed and Wang, Taozhi and Zhang, Juran and Zheng, Yujia and Doyle, Aidan and Sridhar, Pragnya and Agarwal, Arav and others},
  booktitle={Proceedings of the 26th Australasian Computing Education Conference},
  pages={114--123},
  year={2024}
}

@inproceedings{elkins2024teachers,
  title={How Teachers Can Use Large Language Models and Bloom’s Taxonomy to Create Educational Quizzes},
  author={Elkins, Sabina and Kochmar, Ekaterina and Cheung, Jackie CK and Serban, Iulian},
  booktitle={Proceedings of the AAAI Conference on Artificial Intelligence},
  volume={38},
  number={21},
  pages={23084--23091},
  year={2024}
}

@misc{toeic,
    title={{TOEIC Official Website}},
    url={https://www.ets.org/toeic.html},
    author={{Educational Testing Service (ETS)}},
    year= retrievedyear,
    lastaccessed = retrieveddate
}

@book{council2001common,
  title={Common European Framework of Reference for Languages: Learning, Teaching, Assessment},
  author={Council of Europe. Council for Cultural Co-operation. Education Committee. Modern Languages Division},
  year={2001},
  publisher={Cambridge University Press}
}

@article{amano2023manifold,
  title={The Manifold Costs of Being a Non-Native English Speaker in Science},
  author={Amano, Tatsuya and Ram{\'\i}rez-Casta{\~n}eda, Valeria and Berdejo-Espinola, Violeta and Borokini, Israel and Chowdhury, Shawan and Golivets, Marina and Gonz{\'a}lez-Trujillo, Juan David and Monta{\~n}o-Centellas, Flavia and Paudel, Kumar and White, Rachel Louise and Others},
  journal={PLoS biology},
  volume={21},
  number={7},
  pages={e3002184},
  year={2023},
  publisher={Public Library of Science}
}

@inproceedings{guo2018non,
  title={Non-Native English Speakers Learning Computer Programming: Barriers, Desires, and Design Opportunities},
  author={Guo, Philip J},
  booktitle={Proceedings of the 2018 CHI Conference on Human Factors in Computing Systems},
  pages={1--14},
  year={2018}
}

@article{kirkpatrick2011internationalization,
  title={Internationalization or Englishization: Medium of Instruction in Today's Universities},
  author={Kirkpatrick, Thomas Andrew},
  year={2011},
  publisher={Centre for Governance and Citizenship, The Hong Kong Institute of Education}
}

@article{veerasamy2014teaching,
  title={Teaching English Based Programming Courses to English Language Learners/Non-Native Speakers of English},
  author={Veerasamy, Ashok Kumar and Shillabeer, Anna},
  journal={International Proceedings of Economics Development and Research},
  volume={70},
  pages={17},
  year={2014},
  publisher={IACSIT Press}
}

@book{dewey2024democracy,
  title={Democracy and Education},
  author={Dewey, John},
  year={2024},
  publisher={Columbia University Press}
}

@book{collins2006cognitive,
  title={Cognitive Ppprenticeship},
  author={Collins, Allan and Kapur, Manu},
  volume={291},
  year={2006},
  publisher={Citeseer}
}

@article{hwang2008criteria,
  title={Criteria, Strategies and Research Issues of Context-Aware Ubiquitous Learning},
  author={Hwang, Gwo-Jen and Tsai, Chin-Chung and Yang, Stephen JH},
  journal={Journal of Educational Technology \& Society},
  volume={11},
  number={2},
  pages={81--91},
  year={2008},
  publisher={JSTOR}
}

@article{lee2022systematic,
  title={A Systematic Review of Context-Aware Technology Use in Foreign Language Learning},
  author={Lee, Sangmin-Michelle},
  journal={Computer Assisted Language Learning},
  volume={35},
  number={3},
  pages={294--318},
  year={2022},
  publisher={Taylor \& Francis}
}

@article{hidi2006four,
  title={The Four-Phase Model of Interest Development},
  author={Hidi, Suzanne and Renninger, K Ann},
  journal={Educational Psychologist},
  volume={41},
  number={2},
  pages={111--127},
  year={2006},
  publisher={Taylor \& Francis}
}

@article{liu2009context,
  title={A Context-Aware Ubiquitous Learning Environment for Language Listening and Speaking},
  author={Liu, T-Y},
  journal={Journal of Computer Assisted Learning},
  volume={25},
  number={6},
  pages={515--527},
  year={2009},
  publisher={Wiley Online Library}
}

@inproceedings{draxler2023relevance,
  title={Relevance, Effort, and Perceived Quality: Language Learners’ Experiences with AI-Generated Contextually Personalized Learning Material},
  author={Draxler, Fiona and Schmidt, Albrecht and Chuang, Lewis L},
  booktitle={Proceedings of the 2023 ACM Designing Interactive Systems Conference},
  pages={2249--2262},
  year={2023}
}

@inproceedings{edge2011micromandarin,
  title={MicroMandarin: Mobile Language Learning in Context},
  author={Edge, Darren and Searle, Elly and Chiu, Kevin and Zhao, Jing and Landay, James A.},
  booktitle={Proceedings of the SIGCHI conference on human factors in computing systems},
  pages={3169--3178},
  year={2011}
}

@article{hautasaari2019vocabura,
  title={Vocabura: A Method for Supporting Second Language Vocabulary Learning While Walking},
  author={Hautasaari, Ari and Hamada, Takeo and Ishiyama, Kuntaro and Fukushima, Shogo},
  journal={Proceedings of the ACM on Interactive, Mobile, Wearable and Ubiquitous Technologies},
  volume={3},
  number={4},
  pages={1--23},
  year={2019},
  publisher={ACM New York, NY, USA}
}

@inproceedings{yamaoka2022experience,
  title={Experience is the Best Teacher: Personalized Vocabulary Building within the Context of Instagram Posts and Sentences from GPT-3},
  author={Yamaoka, Kanta and Watanabe, Ko and Kise, Koichi and Dengel, Andreas and Ishimaru, Shoya},
  booktitle={Adjunct Proceedings of the 2022 ACM International Joint Conference on Pervasive and Ubiquitous Computing and the 2022 ACM International Symposium on Wearable Computers},
  pages={313--316},
  year={2022}
}

@article{ogata2014ubiquitous,
  title={Ubiquitous Learning Project using Life-Logging Technology in Japan},
  author={Ogata, Hiroaki and Hou, Bin and Li, Mengmeng and Uosaki, Noriko and Mouri, Kosuke and Liu, Songran},
  journal={Journal of Educational Technology \& Society},
  volume={17},
  number={2},
  pages={85--100},
  year={2014},
  publisher={JSTOR}
}

@article{ibrahim2018arbis,
  title={Arbis Pictus: A Study of Vocabulary Learning with Augmented Reality},
  author={Ibrahim, Adam and Huynh, Brandon and Downey, Jonathan and H{\"o}llerer, Tobias and Chun, Dorothy and O'donovan, John},
  journal={IEEE Transactions on Visualization and Computer Graphics},
  volume={24},
  number={11},
  pages={2867--2874},
  year={2018},
  publisher={IEEE}
}

@inproceedings{coleman2012twasebook,
  title={Twasebook: a "Crowdsourced Phrasebook" for Language Learners using Twitter},
  author={Coleman, Graeme W and Hine, Nick A},
  booktitle={Proceedings of the 7th Nordic Conference on Human-Computer Interaction: Making Sense Through Design},
  pages={805--806},
  year={2012}
}

@inproceedings{ding2025unknown,
  title={Unknown Word Detection for English as a Second Language (ESL) Learners using Gaze and Pre-trained Language Models},
  author={Ding, Jiexin and Zhao, Bowen and Wang, Yuntao and Liu, Xinyun and Hao, Rui and Chatterjee, Ishan and Shi, Yuanchun},
  booktitle={Proceedings of the 2025 CHI Conference on Human Factors in Computing Systems},
  pages={1--16},
  year={2025}
}

@article{xu2024device,
  title={On-Device Language Models: A Comprehensive Review},
  author={Xu, Jiajun and Li, Zhiyuan and Chen, Wei and Wang, Qun and Gao, Xin and Cai, Qi and Ling, Ziyuan},
  journal={arXiv preprint arXiv:2409.00088},
  year={2024}
}

@misc{mohamed2025impactllmassistantssoftwaredeveloper,
      title={The Impact of LLM-Assistants on Software Developer Productivity: A Systematic Literature Review}, 
      author={Amr Mohamed and Maram Assi and Mariam Guizani},
      year={2025},
      eprint={2507.03156},
      archivePrefix={arXiv},
      primaryClass={cs.SE},
      url={https://arxiv.org/abs/2507.03156}, 
}

@article{schmitt2014reassessment,
  title={A Reassessment of Frequency and Vocabulary Size in L2 Vocabulary Teaching1},
  author={Schmitt, Norbert and Schmitt, Diane},
  journal={Language Teaching},
  volume={47},
  number={4},
  pages={484--503},
  year={2014},
  publisher={Cambridge University Press}
}

@article{Goetze2022Is,title={Is learning really just believing? A Meta-Analysis of Self-Efficacy and Achievement in SLA},author={Julia Goetze and M. Driver},journal={Studies in Second Language Learning and Teaching},year={2022},doi={10.14746/ssllt.2022.12.2.4}}

@article{Sun2021Relationship,title={Relationship between Second Language English Writing Self-Efficacy and Achievement: A Meta-Regression Analysis},author={Ting Sun and Chuang Wang and R. Lambert and Lan Liu},journal={Journal of Second Language Writing},year={2021},volume={53},pages={100817},doi={10.1016/j.jslw.2021.100817}}

@article{cropley2015relationship,
  title={The Relationship between Work-Related Rumination and Evening and Morning Salivary Cortisol Secretion},
  author={Cropley, Mark and Rydstedt, Leif W. and Devereux, Jason J. and Middleton, Benita},
  journal={Stress and Health},
  volume={31},
  number={2},
  pages={150--157},
  year={2015},
  publisher={Wiley Online Library}
}

@article{sonnentag2015recovery,
  title={Recovery from Job Stress: The Stressor-Detachment Model as an Integrative Framework},
  author={Sonnentag, Sabine and Fritz, Charlotte},
  journal={Journal of Organizational Behavior},
  volume={36},
  number={S1},
  pages={S72--S103},
  year={2015},
  publisher={Wiley Online Library}
}

@article{binnewies2010recovery,
  title={Recovery during the Weekend and Fluctuations in Weekly Job Performance: A Week-Level Study Examining Intra-Individual Relationships},
  author={Binnewies, Carmen and Sonnentag, Sabine and Mojza, Eva J.},
  journal={Journal of Occupational and Organizational Psychology},
  volume={83},
  number={2},
  pages={419--441},
  year={2010},
  publisher={Wiley Online Library}
}

@ARTICLE{11032085,
  author={Sharshar, Ahmed and Khan, Latif U. and Ullah, Waseem and Guizani, Mohsen},
  journal={IEEE Internet of Things Journal}, 
  title={Vision-Language Models for Edge Networks: A Comprehensive Survey}, 
  year={2025},
  volume={12},
  number={16},
  pages={32701-32724},
  doi={10.1109/JIOT.2025.3579032}}

@article{bjork2015desirable,
  title={Desirable Difficulties in Vocabulary Learning},
  author={Bjork, Robert A. and Kroll, Judith F.},
  journal={The American Journal of Psychology},
  volume={128},
  number={2},
  pages={241--252},
  year={2015},
  publisher={University of Illinois Press}
}

@article{bjork2011making,
  title={Making Things Hard on Yourself, but in a Good Way: Creating Desirable Difficulties to Enhance Learning},
  author={Bjork, Elizabeth L. and Bjork, Robert A. and Others},
  journal={Psychology and the Real World: Essays Illustrating Fundamental Contributions to Society},
  volume={2},
  number={59-68},
  pages={56--64},
  year={2011}
}

@article{schmidt1992new,
  title={New Conceptualizations of Practice: Common Principles in Three Paradigms Suggest New Concepts for Training},
  author={Schmidt, Richard A. and Bjork, Robert A.},
  journal={Psychological Science},
  volume={3},
  number={4},
  pages={207--218},
  year={1992},
  publisher={SAGE Publications Sage CA: Los Angeles, CA}
}

@inproceedings{10.1145/3544548.3580817,
author = {Liu, Michael Xieyang and Sarkar, Advait and Negreanu, Carina and Zorn, Benjamin and Williams, Jack and Toronto, Neil and Gordon, Andrew D.},
title = {“What It Wants Me To Say”: Bridging the Abstraction Gap Between End-User Programmers and Code-Generating Large Language Models},
year = {2023},
isbn = {9781450394215},
publisher = {Association for Computing Machinery},
address = {New York, NY, USA},
url = {https://doi.org/10.1145/3544548.3580817},
doi = {10.1145/3544548.3580817},
booktitle = {Proceedings of the 2023 CHI Conference on Human Factors in Computing Systems},
articleno = {598},
numpages = {31},
location = {Hamburg, Germany},
series = {CHI '23}
}

@inproceedings{10.1145/3613904.3642754,
author = {Subramonyam, Hari and Pea, Roy and Pondoc, Christopher and Agrawala, Maneesh and Seifert, Colleen},
title = {Bridging the Gulf of Envisioning: Cognitive Challenges in Prompt Based Interactions with LLMs},
year = {2024},
isbn = {9798400703300},
publisher = {Association for Computing Machinery},
address = {New York, NY, USA},
url = {https://doi.org/10.1145/3613904.3642754},
doi = {10.1145/3613904.3642754},
booktitle = {Proceedings of the 2024 CHI Conference on Human Factors in Computing Systems},
articleno = {1039},
numpages = {19},
location = {Honolulu, HI, USA},
series = {CHI '24}
}

@inproceedings{10.1145/3654777.3676388,
author = {Yen, Ryan and Zhao, Jian},
title = {Memolet: Reifying the Reuse of User-AI Conversational Memories},
year = {2024},
isbn = {9798400706288},
publisher = {Association for Computing Machinery},
address = {New York, NY, USA},
url = {https://doi.org/10.1145/3654777.3676388},
doi = {10.1145/3654777.3676388},
booktitle = {Proceedings of the 37th Annual ACM Symposium on User Interface Software and Technology},
articleno = {58},
numpages = {22},
location = {Pittsburgh, PA, USA},
series = {UIST '24}
}

@article{liu-etal-2024-lost,
    title = "Lost in the Middle: How Language Models Use Long Contexts",
    author = "Liu, Nelson F.  and
      Lin, Kevin  and
      Hewitt, John  and
      Paranjape, Ashwin  and
      Bevilacqua, Michele  and
      Petroni, Fabio  and
      Liang, Percy",
    journal = "Transactions of the Association for Computational Linguistics",
    volume = "12",
    year = "2024",
    address = "Cambridge, MA",
    publisher = "MIT Press",
    url = "https://aclanthology.org/2024.tacl-1.9/",
    doi = "10.1162/tacl_a_00638",
    pages = "157--173"
}

@article{bae2022keep,
  title={Keep Me Updated! Memory Management in Long-Term Conversations},
  author={Bae, Sanghwan and Kwak, Donghyun and Kang, Soyoung and Lee, Min Young and Kim, Sungdong and Jeong, Yuin and Kim, Hyeri and Lee, Sang-Woo and Park, Woomyoung and Sung, Nako},
  journal={arXiv preprint arXiv:2210.08750},
  year={2022}
}

@inproceedings{lungu2018we,
  title={As We May Study: Towards the Web as a Personalized Language Textbook},
  author={Lungu, Mircea F. and van den Brand, Luc and Chirtoaca, Dan and Avagyan, Martin},
  booktitle={Proceedings of the 2018 CHI Conference on Human Factors in Computing Systems},
  pages={1--12},
  year={2018}
}

@inproceedings{kim2025design,
  title={Design Opportunities for Explainable AI Paraphrasing Tools: A User Study with Non-native English Speakers},
  author={Kim, Yewon and Le, Thanh-Long V. and Kim, Donghwi and Lee, Mina and Lee, Sung-Ju},
  booktitle={Proceedings of the 2025 ACM Designing Interactive Systems Conference},
  pages={1061--1083},
  year={2025}
}

@inproceedings{chen2024worddecipher,
  title={WordDecipher: Enhancing Digital Workspace Communication with Explainable AI for Non-native English Speakers},
  author={Chen, Yuexi and Liu, Zhicheng},
  booktitle={Proceedings of the Third Workshop on Intelligent and Interactive Writing Assistants},
  pages={7--10},
  year={2024}
}

@inproceedings{kwon2025better,
  title={How to Better Translate Participant Quotes Using LLMs: Exploring Practices and Challenges of Non-Native English Researchers},
  author={Kwon, Huisung and Min, Soyeong and Lee, Sangsu},
  booktitle={Proceedings of the Extended Abstracts of the CHI Conference on Human Factors in Computing Systems},
  pages={1--11},
  year={2025}
}

@inproceedings{dang2025corpusstudio,
  title={CorpusStudio: Surfacing Emergent Patterns In A Corpus Of Prior Work While Writing},
  author={Dang, Hai and Swoopes, Chelse and Buschek, Daniel and Glassman, Elena L.},
  booktitle={Proceedings of the 2025 CHI Conference on Human Factors in Computing Systems},
  pages={1--19},
  year={2025}
}

@inproceedings{buschek2021impact,
  title={The Impact of Multiple Parallel Phrase Suggestions on Email Input and Composition Behaviour of Native and Non-Native English Writers},
  author={Buschek, Daniel and Z{\"u}rn, Martin and Eiband, Malin},
  booktitle={Proceedings of the 2021 CHI Conference on Human Factors in Computing Systems},
  pages={1--13},
  year={2021}
}

@inproceedings{ito2023use,
  title={Use of an AI-Powered Rewriting Support Software in Context with Other Tools: a Study of Non-Native English Speakers},
  author={Ito, Takumi and Yamashita, Naomi and Kuribayashi, Tatsuki and Hidaka, Masatoshi and Suzuki, Jun and Gao, Ge and Jamieson, Jack and Inui, Kentaro},
  booktitle={Proceedings of the 36th Annual ACM Symposium on User Interface Software and Technology},
  pages={1--13},
  year={2023}
}

@misc{chan_ai_2023,
	title = {The {AI} generation gap: {Are} {Gen} {Z} students more interested in adopting generative {AI} such as {ChatGPT} in teaching and learning than their {Gen} {X} and {Millennial} {Generation} teachers?},
	shorttitle = {The {AI} generation gap},
	url = {http://arxiv.org/abs/2305.02878},
	doi = {10.48550/arXiv.2305.02878},
	urldate = {2023-10-19},
	publisher = {arXiv},
	author = {Chan, Cecilia Ka Yuk and Lee, Katherine K. W.},
	month = may,
	year = {2023},
	note = {arXiv:2305.02878 [cs]},
}

@inproceedings{10.1145/3706598.3713778,
author = {Lee, Hao-Ping (Hank) and Sarkar, Advait and Tankelevitch, Lev and Drosos, Ian and Rintel, Sean and Banks, Richard and Wilson, Nicholas},
title = {The Impact of Generative AI on Critical Thinking: Self-Reported Reductions in Cognitive Effort and Confidence Effects From a Survey of Knowledge Workers},
year = {2025},
isbn = {9798400713941},
publisher = {Association for Computing Machinery},
address = {New York, NY, USA},
url = {https://doi.org/10.1145/3706598.3713778},
doi = {10.1145/3706598.3713778},
booktitle = {Proceedings of the 2025 CHI Conference on Human Factors in Computing Systems},
articleno = {1121},
numpages = {22},
series = {CHI '25}
}

@inproceedings{higasa2024keep,
  title={Keep Eyes on the Sentence: An Interactive Sentence Simplification System for English Learners Based on Eye Tracking and Large Language Models},
  author={Higasa, Taichi and Tanaka, Keitaro and Feng, Qi and Morishima, Shigeo},
  booktitle={Extended Abstracts of the CHI Conference on Human Factors in Computing Systems},
  pages={1--7},
  year={2024}
}

@book{anthony2018introducing,
  title={Introducing English for Specific Purposes},
  author={Anthony, Laurence},
  year={2018},
  publisher={Routledge}
}

@article{gilmore2007authentic,
  title={Authentic Materials and Authenticity in Foreign Language Learning},
  author={Gilmore, Alex},
  journal={Language Teaching},
  volume={40},
  number={2},
  pages={97--118},
  year={2007},
  publisher={Cambridge University Press}
}

@article{bielousova2017developing,
  title={Developing Materials for English for Specific Purposes Online Course within the Blended Learning Concept},
  author={Bielousova, Rimma},
  journal={Tem Journal},
  volume={6},
  number={3},
  pages={637--642},
  year={2017},
  publisher={UIKTEN-Association for Information Communication Technology Education}
}

@article{benavent2011use,
  title={Use of Authentic Materials in the ESP Classroom},
  author={Benavent, Gabriela Torregrosa and Pe{\~n}amar{\'\i}a, Sonsoles S{\'a}nchez-Reyes},
  journal={Online Submission},
  volume={20},
  pages={89--94},
  year={2011},
  publisher={ERIC}
}

@inproceedings{de2002visible,
  title={Visible or Invisible Links?},
  author={De Ridder, Isabelle},
  booktitle={CHI'02 Extended Abstracts on Human Factors in Computing Systems},
  pages={624--625},
  year={2002}
}

@article{sanko2006effects,
  title={The Effects of Hypertextual Input Modification on L2 Vocabulary Acquisition and Retention},
  author={Sank{\'o}, Gyula},
  journal={University of P{\'e}cs Roundtable 2006: Empirical Studies in English Applied Linguistics},
  pages={157},
  year={2006}
}

@inproceedings{azab2013nlp,
  title={An NLP-Based Reading Tool for Aiding Non-Native English Readers},
  author={Azab, Mahmoud and Salama, Ahmed and Oflazer, Kemal and Shima, Hideki and Araki, Jun and Mitamura, Teruko},
  booktitle={Proceedings of the International Conference Recent Advances in Natural Language Processing RANLP 2013},
  pages={41--48},
  year={2013}
}

@article{rowland2014effect,
  title={The Effect of Testing Versus Restudy on Retention: a Meta-Analytic Review of the Testing Effect},
  author={Rowland, Christopher A.},
  journal={Psychological Bulletin},
  volume={140},
  number={6},
  pages={1432},
  year={2014},
  publisher={American Psychological Association}
}

@article{thompson2019practice,
  title={Practice Makes Perfect? A Review of Second Language Teaching Methods},
  author={Thompson, Colin},
  journal={The Bulletin of the Graduate School of Josai International University},
  volume={22},
  number={55-69},
  year={2019}
}

@book{johnson1997language,
  title={Language Teaching and Skill Learning},
  author={Johnson, Keith},
  year={1997},
  publisher={Blackwell Oxford, England}
}

@article{toppino2018metacognitive,
  title={Metacognitive Control in Self-Regulated Learning: Conditions Affecting the Choice of Restudying Versus Retrieval Practice},
  author={Toppino, Thomas C. and LaVan, Melissa H. and Iaconelli, Ryan T.},
  journal={Memory \& Cognition},
  volume={46},
  number={7},
  pages={1164--1177},
  year={2018},
  publisher={Springer}
}

@inproceedings{draxler2022agenda,
  title={Agenda-and Activity-Based Triggers for Microlearning},
  author={Draxler, Fiona and Brenner, Julia Maria and Eska, Manuela and Schmidt, Albrecht and Chuang, Lewis L},
  booktitle={Proceedings of the 27th International Conference on Intelligent User Interfaces},
  pages={620--632},
  year={2022}
}

@article{gassler2004integrated,
  title={Integrated Micro Learning--An Outline of the Basic Method and First Results},
  author={Gassler, Gerhard and Hug, Theo and Glahn, Christian},
  journal={Interactive Computer Aided Learning},
  volume={4},
  pages={1--7},
  year={2004},
  publisher={Kassel Univeristy Press}
}

@article{devellis2006classical,
  title={Classical Test Theory},
  author={DeVellis, Robert F.},
  journal={Medical care},
  volume={44},
  number={11},
  pages={S50--S59},
  year={2006},
  publisher={LWW}
}

@article{kuhlthau1999role,
  title={The Role of Experience in the Information Search Process of an Early Career Information worker: Perceptions of uncertainty, complexity, construction, and sources},
  author={Kuhlthau, Carol Collier},
  journal={Journal of the American Society for Information Science},
  volume={50},
  number={5},
  pages={399--412},
  year={1999},
  publisher={Wiley Online Library}
}

@inproceedings{culbertson2016social,
  title={Social Situational Language Learning through an Online 3D Game},
  author={Culbertson, Gabriel and Wang, Shiyu and Jung, Malte and Andersen, Erik},
  booktitle={Proceedings of the 2016 CHI Conference on Human Factors in Computing Systems},
  pages={957--968},
  year={2016}
}

@inproceedings{edge2012memreflex,
  title={MemReflex: Adaptive Flashcards for Mobile Microlearning},
  author={Edge, Darren and Fitchett, Stephen and Whitney, Michael and Landay, James},
  booktitle={Proceedings of the 14th International Conference on Human-Computer Interaction with Mobile Devices and Services},
  pages={431--440},
  year={2012}
}

@inproceedings{inie2021aiki,
  title={Aiki-Turning Online Procrastination into Microlearning},
  author={Inie, Nanna and Lungu, Mircea F.},
  booktitle={Proceedings of the 2021 CHI Conference on Human Factors in Computing Systems},
  pages={1--13},
  year={2021}
}

@inproceedings{kim2024interrupting,
  title={Interrupting for Microlearning: Understanding Perceptions and Interruptibility of Proactive Conversational Microlearning Services},
  author={Kim, Minyeong and Lee, Jiwook and Koh, Youngji and Lee, Chanhee and Lee, Uichin and Kim, Auk},
  booktitle={Proceedings of the 2024 CHI Conference on Human Factors in Computing Systems},
  pages={1--21},
  year={2024}
}

@article{rawson2011optimizing,
  title={Optimizing Schedules of Retrieval Practice for Durable and Efficient Learning: How Much Is Enough?},
  author={Rawson, Katherine A. and Dunlosky, John},
  journal={Journal of Experimental Psychology: General},
  volume={140},
  number={3},
  pages={283},
  year={2011},
  publisher={American Psychological Association}
}

@article{robinson1995attention,
  title={Attention, Memory, and the “Noticing” Hypothesis},
  author={Robinson, Peter},
  journal={Language Learning},
  volume={45},
  number={2},
  pages={283--331},
  year={1995},
  publisher={Wiley Online Library}
}

@inproceedings{arakawa2022vocabencounter,
  title={VocabEncounter: NMT-Powered Vocabulary Learning by Presenting Computer-Generated Usages of Foreign Words into Users’ Daily Lives},
  author={Arakawa, Riku and Yakura, Hiromu and Kobayashi, Sosuke},
  booktitle={Proceedings of the 2022 CHI Conference on Human Factors in Computing Systems},
  pages={1--21},
  year={2022}
}

@inproceedings{peng2023storyfier,
  title={Storyfier: Exploring Vocabulary Learning Support with Text Generation Models},
  author={Peng, Zhenhui and Wang, Xingbo and Han, Qiushi and Zhu, Junkai and Ma, Xiaojuan and Qu, Huamin},
  booktitle={Proceedings of the 36th Annual ACM Symposium on User Interface Software and Technology},
  pages={1--16},
  year={2023}
}

@article{yamada2012development,
  title={Development and Evaluation of English Listening Study Materials for Business People Who Use Mobile Devices},
  author={Yamada, Masanori and Kitamura, Satoshi and Shimada, Noriko and Utashiro, Takafumi and Shigeta, Katsusuke and Yamaguchi, Etsuji and Harrison, Richard and Yamauchi, Yuhei},
  journal={Calico Journal},
  volume={29},
  number={1},
  pages={44--66},
  year={2012},
  publisher={Equinox Publishing Ltd.}
}

@article{hanusz2016shapiro,
  title={Shapiro--Wilk Test with Known Mean},
  author={Hanusz, Zofia and Tarasinska, Joanna and Zielinski, Wojciech},
  journal={REVSTAT-Statistical Journal},
  volume={14},
  number={1},
  pages={89--100},
  year={2016}
}

@article{ebbinghaus2013image,
  title={Memory: A Contribution to Experimental Psychology},
  author={Ebbinghaus, Hermann},
  journal={Annals of neurosciences},
  volume={20},
  number={4},
  pages={155},
  year={2013}
}

@inproceedings{leong2024putting,
  title={Putting Things into Context: Generative AI-Enabled Context Personalization for Vocabulary Learning Improves Learning Motivation},
  author={Leong, Joanne and Pataranutaporn, Pat and Danry, Valdemar and Perteneder, Florian and Mao, Yaoli and Maes, Pattie},
  booktitle={Proceedings of the 2024 CHI Conference on Human Factors in Computing Systems},
  pages={1--15},
  year={2024}
}

@inproceedings{wang2019minddot,
  title={MindDot: Supporting Effective Cognitive Behaviors in Concept Map-Based Learning Environments},
  author={Wang, Shang and Sonmez Unal, Deniz and Walker, Erin},
  booktitle={Proceedings of the 2019 CHI Conference on Human Factors in Computing Systems},
  pages={1--14},
  year={2019}
}

@inproceedings{kovacs2015feedlearn,
  title={FeedLearn: Using Facebook Feeds for Microlearning},
  author={Kovacs, Geza},
  booktitle={Proceedings of the 33rd Annual ACM Conference Extended Abstracts on Human Factors in Computing Systems},
  pages={1461--1466},
  year={2015}
}

@article{epstein2022piggyback,
author = {Epstein, Daniel A. and Liu, Fannie and Monroy-Hern\'{a}ndez, Andr\'{e}s and Wang, Dennis},
title = {Revisiting Piggyback Prototyping: Examining Benefits and Tradeoffs in Extending Existing Social Computing Systems},
year = {2022},
issue_date = {November 2022},
publisher = {Association for Computing Machinery},
address = {New York, NY, USA},
volume = {6},
number = {CSCW2},
url = {https://doi.org/10.1145/3555557},
doi = {10.1145/3555557},
journal = {Proc. ACM Hum.-Comput. Interact.},
month = nov,
articleno = {456},
numpages = {28}
}

@inproceedings{grevet2015piggyback,
author = {Grevet, Catherine and Gilbert, Eric},
title = {Piggyback Prototyping: Using Existing, Large-Scale Social Computing Systems to Prototype New Ones},
year = {2015},
isbn = {9781450331456},
publisher = {Association for Computing Machinery},
address = {New York, NY, USA},
url = {https://doi.org/10.1145/2702123.2702395},
doi = {10.1145/2702123.2702395},
booktitle = {Proceedings of the 33rd Annual ACM Conference on Human Factors in Computing Systems},
pages = {4047–4056},
numpages = {10},
location = {Seoul, Republic of Korea},
series = {CHI '15}
}
